\documentclass[%
  reprint,
prb,
]{revtex4-1}

\usepackage{array}
\usepackage{calc}
\usepackage{etoolbox,siunitx}
\robustify\bfseries
\usepackage{grffile}
\usepackage{graphicx}
\usepackage{color}
\usepackage{dcolumn}
\usepackage{bm}
\usepackage{multirow}
\usepackage{pbox}
\usepackage{tabularx,booktabs}
\usepackage{amsmath}
\usepackage{subfigure}
\usepackage{physics}
\usepackage{longtable}
\usepackage{soul}

\definecolor{pink}{rgb}{0.858, 0.188, 0.478}
\definecolor{green}{rgb}{0, 0.4, 0}

\newcolumntype{P}[1]{>{\centering\arraybackslash}p{#1}}
\newcolumntype{M}[1]{>{\centering\arraybackslash}m{#1}}
\soulregister\cite7
\soulregister\ref7

\makeatletter
\def\NAT@def@citea{\def\@citea{\NAT@separator}}
\makeatother

\begin{document}
\title{Non-equilibrium Green's function predictions of band tails and band gap narrowing in III-V semiconductors and nanodevices}

\author{Prasad Sarangapani}
\affiliation{School of Electrical and Computer Engineering, Purdue University, West Lafayette, IN 47906, USA}
\author{Yuanchen Chu}
\affiliation{School of Electrical and Computer Engineering, Purdue University, West Lafayette, IN 47906, USA}
\author{James Charles}
\affiliation{School of Electrical and Computer Engineering, Purdue University, West Lafayette, IN 47906, USA}
\author{Tillmann Kubis}
\affiliation{School of Electrical and Computer Engineering, Purdue University, West Lafayette, IN 47906, USA}
\affiliation{Network for Computational Nanotechnology, Purdue University, West Lafayette, IN 47906, USA}
\affiliation{Purdue Center for Predictive Materials and Devices, Purdue University,  West Lafayette, IN 47906, USA}

\email{psaranga@purdue.edu}

\begin{abstract} 
High-doping induced Urbach tails and band gap narrowing play a significant role in determining the performance of tunneling devices and optoelectronic devices such as tunnel field-effect transistors (TFETs), Esaki diodes and light-emitting diodes. 
In this work, Urbach tails and band gap narrowing values are calculated explicitly for GaAs, InAs, GaSb and GaN as well as ultra-thin bodies and nanowires of the same.
Electrons are solved in the non-equilibrium Green's function method in multi-band atomistic tight binding. 
Scattering on polar optical phonons and charged impurities is solved in the self-consistent Born approximation. 
The corresponding nonlocal scattering self-energies as well as their numerically efficient formulations are introduced for ultra-thin bodies and nanowires.
Predicted Urbach band tails and conduction band gap narrowing agree well with  experimental literature for a range of temperatures and doping concentrations.
Polynomial fits of the Urbach tail and band gap narrowing as a function of doping are tabulated for quick reference.
\end{abstract}

\maketitle

\section{Introduction}
The need for ultra-low power applications, efficient lighting and renewable energy sources have resulted in the development of novel devices such as the tunnel field-effect transistors (TFETs)~\cite{lind2015iii, avci2015tunnel, seabaugh2010low}, GaN/InGaN light-emitting diodes (LEDs)~\cite{geng2018quantitative, laubsch2010high, guo2010catalyst} and high-performance solar cells~\cite{krogstrup2013single, bailey2011near, wallentin2013inp}. 
Carrier transport and sub-60 mV subthreshold slope (SS) performance in TFETs, optical recombination and carrier generation in LEDs is highly dependent on a good description of conduction and valence band properties. 
Tailing of band edge states (known as Urbach tails/band tails) and band gap narrowing can significantly alter the device behaviour. 
For instance, the switching behaviour of TFETs is drastically affected by such band tailing~\cite{agarwal2014band, bizindavyi2018band}. 
Exponentially decaying band tail states (below the conduction band and above the valence band) are known to fundamentally limit the lowest achievable SS in TFETs~\cite{lu2014tunnel, agarwal2014band}. 
On the other hand, band gap narrowing is known to alter the optical frequency at which recombination and carrier generation occurs in LEDs and solar cells. 
It also shifts the turn-on and threshold voltage of optical devices and the tunneling current in TFETs~\cite{geng2018quantitative,oehme2013direct}.
	
Band tailing and band gap narrowing effects are mainly attributed to the interaction of electrons and holes with phonons, randomly distributed dopant impurity atoms and native lattice disorders and defects. 
They exhibit a strong dependence on temperature and doping concentration.~\cite{halperin1966impurity, halperin1967impurity, john1986theory,jain1991simple}
Though the effect has been studied for quite some time, actual values for specific materials are based either on heuristic models or parameters that are directly extracted from experimental observations.~\cite{halperin1966impurity, halperin1967impurity, jain1991simple, zhang2016effect, bizindavyi2018band}
This situation is inconvenient, since the validity of the heuristic expressions is limited by the doping range and underlying assumptions on the band dispersion. 
In addition, band gaps and tailing of confined devices such as ultra-thin body and nanowires are notoriously hard to predict without the presence of available experimental data. 
	
In this work, band-tailing and band gap narrowing are predicted with scattering self-energies in the framework of the non-equilibrium Green's function (NEGF) method. 
The predicted values depend  only on material dependent parameters. 
The NEGF method is among the most detailed quantum transport methods for electronic, thermal and optoelectronic effects in a variety of nanodevices ~\cite{datta2000nanoscale, lake1997single, markussen2009electron, lee2002nonequilibrium, steiger2009modelling, kubis2009theory}. 
It has been applied in modeling transistors \cite{luisier2009atomistic, afzalian2018physics, ameen2017combination}, resonant-tunneling devices \cite{bowen1997quantitative}, metal-semiconductor contacts \cite{sarangapani2018atomistic, hegde2014effect}, phonon transport across interfaces \cite{miao2016buttiker}, GaN/InGaN light-emitting diodes \cite{geng2018quantitative} with quantitative agreements with experimental data.
In this work, incoherent scattering is modeled through scattering self-energies within the self-consistent Born approximation (SCBA).~\cite{lake1997single}
Amongst the scattering mechanisms present in doped III-V semiconductors, polar optical phonons (POP) and charged impurity scattering mechanisms are the dominant mechanisms~\cite{riddoch1983scattering,fischetti1988monte}.
Both scattering mechanisms are considered including electrostatic screening effects. 
The scattering self-energies for both scattering mechanisms are derived for ultra-thin bodies and nanowires and then verified against Fermi's golden rule. 
Urbach tails and band gap narrowing parameters are extracted from the NEGF predicted density of states (DoS) for GaAs, InAs, GaSb and GaN and respective nanodevices. 
Simulated results are benchmarked against experimental data. 
A simple analytical formula for Urbach tail and band gap narrowing in ultra-thin bodies and nanowire devices is fit against the NEGF predictions and respective parameters are provided as guideline values.

\section{Simulation approach}
\subsection{Device representation}
All NEGF simulations in this work have been performed with the multi-purpose nanodevice simulation tool NEMO5.~\cite{steiger2011nemo5}
Bulk, ultra-thin body and nanowire devices considered in this work are all modeled in atomic resolution with the atoms in their native lattice. 
GaAs, InAs and GaSb are considered in the conventional zincblende cystal structure and GaN in the wurtzite structure. 
A 10-band sp3d5s* tight-binding Hamiltonian is used to describe conduction and valence bands.~\cite{klimeck2000sp3s,jancu2002transferable} 
Both POP and charged impurity scattering mechanisms are long-ranged and the extent of non-locality is determined by the electrostatic screening length. 
For all the discussions in the subsequent sections, the respective device is assumed to be in equilibrium with the density corresponding to the doping concentration. 
Self-energies derived are solved self-consistently with the corresponding Green's function following the self-consistent Born approximation scheme. 
The convergence is achieved once maximum relative particle current variation of $10^{-5}$ is given throughout the device. 
The retarded ($G^{R}$) and lesser than ($G^{<}$) Green's functions are solved with
\begin{equation}
\begin{array}{c}
G^{R}=\left(EI-H-\Sigma_{pop}^{R}-\Sigma_{imp}^{R}-\Sigma_{source}^{R}-\Sigma_{drain}^{R}\right)^{-1}\\
\\
G^{<}=G^{R}\left(\Sigma_{pop}^{<}+\Sigma_{imp}^{<}+\Sigma_{source}^{<}+\Sigma_{drain}^{<}\right)G^{R\dagger}
\end{array}
\end{equation}
where $\Sigma_{pop}^R$ and $\Sigma_{pop}^<$ are the retarded and lesser scattering self-energies for polar optical phonon scattering and $\Sigma_{imp}^R$ and $\Sigma_{imp}^<$ are the retarded and lesser scattering self-energies for charged impurity scattering.

\subsection{Scattering self-energy formulas}
\label{Sec:selfenergies}

Charged impurity scattering is modeled assuming homogeneously distributed impurity atoms. 
In this work, Brooks-Herring impurity scattering~\cite{brooks1951scattering} is employed due to the considered doping concentration ranging from $1\times10^{17}\;cm^{-3}$ up to $5\times10^{19}\;cm^{-3}$). 
The screened impurity scattering potential is given by 
    \begin{equation}
      \left|U_{imp,q}\right|^{2}=\dfrac{Z^{2}e^{4}}{\epsilon_{0}^2\epsilon_{s}^2\left(q^{2}+\zeta^{-2}\right)^2}
    \label{Eq:Uq_imp}
    \end{equation}
where $Z$ is the number of electrons in a single dopant atom (taken to be 1 in this work) and $\zeta$ is the screening length. 
The positive elementary charge is given by $e$ and $\epsilon_{0}$ is the vaccuum permitivity.
Electrostatic screening of conduction band electrons is calculated within the Lindhard formalism~\cite{lindhard1954properties} where the screening length is represented as 
\begin{equation}
\begin{array}{c}
\zeta_{Lindhard}=\left(\dfrac{e^{2}}{\epsilon_{0}\epsilon_{s}}\dfrac{-2}{(2\pi)^{3}}{\displaystyle \int}d\vec{q}\left.\dfrac{\partial f}{\partial\epsilon}\right|_{\epsilon(\vec{q})}\right)^{-1/2}
\end{array}
\label{eq:Lindhard}
\end{equation}
where $f$ is the electronic distribution function and the momentum integral runs over the first Brillouin zone. For this and all following expressions, three dimensional vectors are denoted with a \hspace{1pt} $\vec{\vphantom{ q }}$ \hspace{1pt} and two dimensional vectors are boldfaced.
 
The Fr\"ohlich coupling is considered for the electron-phonon interaction potential~\cite{frohlich1952interaction}
\begin{equation}
\left|U_{pop,q}\right|^{2}=e^{2}\dfrac{\hbar\omega_{LO}}{2\epsilon_{0}}\left(\dfrac{1}{\epsilon_{\infty}}-\dfrac{1}{\epsilon_{s}}\right)\dfrac{q^{2}}{(q^{2}+\zeta^{-2})^{2}}
\label{Eq:Uq_pop}
\end{equation}
The static and dynamic dielectric constants are represented by $\epsilon_s$ and $\epsilon_\infty$ respectively. 
The phonon frequency and momentum are denoted with $\hbar\omega_{LO}$ and $q$, respectively. 
The LO phonons are assumed as plane waves in all directions for bulk, UTBs and nanowires.
The longitudinal-optical (LO) phonon frequency is assumed to be momentum independent and to agree with its value at the $\Gamma$ point. 
Phonon frequencies and dielectric constants used for GaAs, InAs, GaSb and GaN are listed in Table~\ref{table:params}. 

\begin{table*}
\begin{centering}
\begin{ruledtabular}
\begin{tabular}{|c|c|c|c|c|}
\hline 
 & \textbf{GaN} & \textbf{GaAs} & \textbf{GaSb} & \textbf{InAs}\tabularnewline
\hline 
LO phonon frequency ($\hbar\omega_{LO}$) & 92 meV & 36 meV & 30 meV & 30 meV\tabularnewline
\hline 
Static dielectric constant ($\epsilon_s$) & 10.4 & 12.95 & 15.69 & 12.3 \tabularnewline
\hline 
Infinite freq. dielectric constant ($\epsilon_\infty$) & 5.47 & 10.9 & 14.4 & 11.6 \tabularnewline
\hline 
\end{tabular}
\end{ruledtabular}
\protect\caption{Material parameters used for GaN, GaAs, GaSb and InAs. GaN material parameters have been obtained from Refs.~\cite{LandoltBornstein2001:sm_lbs_978-3-540-31355-7_87,LandoltBornstein2011:sm_lbs_978-3-642-14148-5_222}, GaAs material parameters from Refs.~\cite{LandoltBornstein2011:sm_lbs_978-3-642-14148-5_102,Villars2016:sm_isp_ppp_5b220e9b7c082d4f92d21887a4be91a0}, GaSb material parameters from Refs.~\cite{LandoltBornstein2001:sm_lbs_978-3-540-31355-7_118,LandoltBornstein2001:sm_lbs_978-3-540-31355-7_124} and InAs parameters from Refs.~\cite{LandoltBornstein2002:sm_lbs_978-3-540-31356-4_356,LandoltBornstein2010:sm_lbs_978-3-540-92140-0_163}}
\label{table:params}
\par\end{centering}
\end{table*}

Electron scattering on charged impurities and polar optical phonons is non-local which yields significant numerical load. 
To limit the numerical burden, the integrals in the scattering self-energies are solved analytically as far as possible. 
Therefore, the self-energies are formulated separately for each confinement setting (ultra-thin body and nanowires). 
For completeness, the formulas for three dimensional, unconfined (bulk) systems are listed here as well.
All scattering self-energy formulas below are given in position representation for nanowires, in position space and one dimensional momentum space for ultra-thin bodies and in position and two dimensional momentum space for bulk systems, respectively. 
The position representation translates directly into the atomistic tight binding, since all atoms have different coordinates. 
Interorbital transitions are only included on the same atom, i.e. when the two position coordinates of the scattering self-energy agree. On the same atom, all interorbital transitions are considered equally likely. 
Nonlocal scattering between different atoms is limited to same orbital types only to reduce the peak memory usage and the time to solution.
To correct this underestimation of nonlocal scattering, the scattering self-energies are multiplied with a compensation factor that is deduced from Fermi's golden rule as detailed in Sec.~\ref{Sec:FGR}.  

The general expression for the retarded and lesser scattering self-energy of electrons scattering on homogeneously distributed charged impurities read~\cite{kubis2009theory}
\begin{equation}
\begin{gathered}
\Sigma^{<,R}_{imp}\left(\vec{x}_{1},\vec{x}_{2},E\right)=\dfrac{1}{(2\pi)^{3}}N_{D}{\displaystyle \int}d\vec{q}\left|U_{imp,q}\right|^{2}e^{i\vec{q}\cdot(\vec{x}_{1}-\vec{x}_{2})}\\
\times G^{<,R}(\vec{x}_{1},\vec{x}_{2},E)
\end{gathered}
\label{eq:impselfenergygeneralexp}
\end{equation}
where the momentum integral runs over the first Brillouin zone. Deriving expressions for different levels of periodicity, we have the following expressions for bulk, ultra-thin body and nanowires. 
Impurity scattering self-energies in 3D periodic systems (bulk) can be expressed as \cite{kubis2009theory, lake1997single}
\begin{equation} \label{eq:implessselfenergy1D}
\begin{gathered}
{\displaystyle \Sigma^{<,R}_{imp}\left(x_{1},x_{2},\boldsymbol{k},E\right)=}\dfrac{N_{D}}{4\left(2\pi\right)^{2}}{\displaystyle \left(\dfrac{e^{2}}{\epsilon_{0}\epsilon_{s}}\right)^{2}}\\
\times{\displaystyle \int \boldsymbol{dk'}}
\left[\left(\dfrac{|x_{1}-x_{2}|+1/\sqrt{|\boldsymbol{k}-\boldsymbol{k'}|^{2}+\zeta^{-2}}}{|\boldsymbol{k}-\boldsymbol{k'}|^{2}+\zeta^{-2}}\right)\right.\\
\left.\times e^{-\left(\sqrt{|\boldsymbol{k}-\boldsymbol{k'}|^{2}+\zeta^{-2}}\left|x_{1}-x_{2}\right|\right)}G^{<,R}\left(x_{1},x_{2},\boldsymbol{k'},E\right)\vphantom{\dfrac{|x_{1}-x_{2}|+1/\sqrt{|\boldsymbol{k}-\boldsymbol{k'}|^{2}+\zeta^{-2}}}{|\boldsymbol{k}-\boldsymbol{k'}|^{2}+\zeta^{-2}}}\right]
\end{gathered}
\end{equation}
Charged impurity scattering self-energies for UTB devices read 
\begin{equation} \label{eq:implessselfenergy2D}
\begin{gathered}
{\displaystyle \Sigma^{<,R}_{imp}\left(\boldsymbol{{x}_{1}},\boldsymbol{{x}_{2}},k,E\right)}=\dfrac{N_{D}}{(8\pi)^{2}}\left(\dfrac{e^{2}}{\epsilon_{0}\epsilon_{s}}\right)^{2}\\
\times\int dk'I(k,k',\boldsymbol{{x}_{1}},\boldsymbol{{x}_{2}})G^{<,R}\left(\boldsymbol{{x}_{1}},\boldsymbol{{x}_{2}},k',E\right)
\end{gathered}
\end{equation}
where
\begin{equation}
\begin{gathered}
I(k,k^{'},\boldsymbol{{x}_{1}},\boldsymbol{{x}_{2}})=\\
\left\{ \begin{gathered}
\dfrac{\pi\left|\boldsymbol{{x}_{1}}-\boldsymbol{{x}_{2}}\right|}{\sqrt{\left(k-k'\right)^{2}+\zeta^{-2}}}\\
\times K_{1}\left(\sqrt{\left(k-k'\right)^{2}+\zeta^{-2}}\left|\boldsymbol{{x}_{1}}-\boldsymbol{{x}_{2}}\right|\right),\;\left|\boldsymbol{{x}_{1}}-\boldsymbol{{x}_{2}}\right|\neq0\\
\dfrac{\pi}{\left(k-k'\right)^{2}+\zeta^{-2}},\;\left|\boldsymbol{{x}_{1}}-\boldsymbol{{x}_{2}}\right|=0
\end{gathered}\right.
\end{gathered}
\end{equation}
where $K_1$ is the Bessel-K function of order 1. Finally, charged impurity scattering self-energies for nanowires can be expressed as 
\begin{equation} \label{eq:implessselfenergy3D}
\begin{gathered}
{\displaystyle \Sigma^{<,R}_{imp}\left(\vec{x}_{1},\vec{x}_{2},E\right)}=\dfrac{N_{D}}{8\pi}\left(\dfrac{e^{2}}{\epsilon_{0}\epsilon_{s}}\right)^{2}\\
\times \zeta e^{-\left|\vec{x}_{1}-\vec{x}_{2}\right|/\zeta}G^{<,R}\left(\vec{x}_{1},\vec{x}_{2},E\right)
\end{gathered}
\end{equation}
All analytical results for the scattering kernels above assume momentum integrations run from $(-\infty,\infty)$ rather than over the first Brillouin zone only. 
Since the scattering potentials decay sharply with increasing momentum, this approximation does not affect the overall result.~\cite{kubis2009theory} 

The general expression for the lesser scattering self-energy of electrons scattering with phonons reads~\cite{wacker2002semiconductor}
\begin{equation}
\begin{gathered}
\Sigma^{<}_{pop}\left(\vec{x}_{1},\vec{x}_{2},E\right)=\dfrac{1}{(2\pi)^{3}}{\displaystyle \int}d\vec{q}\left|U_{pop,q}\right|^{2}e^{i\vec{q}\cdot\left(\vec{x}_{1}-\vec{x}_{2}\right)} \\
\times\left[\vphantom{\dfrac{1}{2}}N_{ph}G^{<}\left(\vec{x}_{1},\vec{x}_{2},E-\hbar\omega_{LO}\right)\right.\\
\left.+(1+N_{ph})G^{<}\left(\vec{x}_{1},\vec{x}_{2},E+\hbar\omega_{LO}\right)\vphantom{\dfrac{1}{2}}\right]
\end{gathered}
\label{eq:generalselfenergyeqn}
\end{equation}
where $N_{ph}$ is the number of phonons due to the Bose distribution evaluated for the LO phonon energy. 
The momentum integral in Eq.~\ref{eq:generalselfenergyeqn} runs over the first Brillouin zone. 
Starting with this expression and solving for each degree of periodicity, we get  
\begin{widetext}    
\begin{equation} \label{eq:popselfenergyless1D}
\begin{gathered}
{\displaystyle \Sigma^{<}_{pop}}\left(x_{1},x_{2},\boldsymbol{k},E\right) = \dfrac{e^2 \pi}{(2\pi)^{3}}\left(\dfrac{1}{\epsilon_{\infty}}-\dfrac{1}{\epsilon_{s}}\right)\dfrac{\hbar\omega_{LO}}{2\epsilon_{0}}\\
\times{\displaystyle \int}\boldsymbol{dk'}\dfrac{e^{-\left(\sqrt{|\boldsymbol{k}-\boldsymbol{k'}|^2+\zeta^{-2}}\left|x_{1}-x_{2}\right|\right)}}{\sqrt{|\boldsymbol{k}-\boldsymbol{k'}|^2+\zeta^{-2}}}\left(1-\dfrac{\zeta^{-2}\left|x_{1}-x_{2}\right|}{2\sqrt{|\boldsymbol{k}-\boldsymbol{k'}|^2+\zeta^{-2}}}-\dfrac{\zeta^{-2}}{2\left(|\boldsymbol{k}-\boldsymbol{k'}|^2+\zeta^{-2}\right)}\right)\\
\times\left[\vphantom{\dfrac{1}{2}}N_{ph}G^{<}\left(x_{1},x_{2},\boldsymbol{k'},E-\hbar\omega_{LO}\right)+\left(N_{ph}+1\right)G^{<}\left(x_{1},x_{2},\boldsymbol{k'},E+\hbar\omega_{LO}\right)\vphantom{\dfrac{1}{2}}\right]
\end{gathered}
\end{equation}
\end{widetext}
\begin{widetext}
\begin{equation}
\begin{gathered}
{\displaystyle \Sigma^{R}_{pop}}\left(x_{1},x_{2},\boldsymbol{k},E\right)
=\dfrac{e^2 \pi}{(2\pi)^{3}}\left(\dfrac{1}{\epsilon_{\infty}}-\dfrac{1}{\epsilon_{s}}\right)\dfrac{\hbar\omega_{LO}}{2\epsilon_{0}}\\
\times{\displaystyle \int}d\boldsymbol{k'}\dfrac{e^{-\left(\sqrt{|\boldsymbol{k}-\boldsymbol{k'}|^2+\zeta^{-2}}\left|x_{1}-x_{2}\right|\right)}}{\sqrt{|\boldsymbol{k}-\boldsymbol{k'}|^2+\zeta^{-2}}}\left(1-\dfrac{\zeta^{-2}\left|x_{1}-x_{2}\right|}{2\sqrt{|\boldsymbol{k}-\boldsymbol{k'}|^2+\zeta^{-2}}}-\dfrac{\zeta^{-2}}{2\left(|\boldsymbol{k}-\boldsymbol{k'}|^2+\zeta^{-2}\right)}\right)\\
\times\left[\vphantom{\dfrac{1}{2}}\left(N_{ph}+1\right)G^{R}\left(x_{1},x_{2},\boldsymbol{k'},E-\hbar\omega_{LO}\right)+N_{ph}G^{R}\left(x_{1},x_{2},\boldsymbol{k'},E+\hbar\omega_{LO}\right) + \dfrac{1}{2}G^{<}\left(x_{1},x_{2},\boldsymbol{k'},E-\hbar\omega_{LO}\right)\right.\\
-\dfrac{1}{2}G^{<}\left(x_{1},x_{2},\boldsymbol{k'},E+\hbar\omega_{LO}\right)
+ \left. i {\displaystyle \int\dfrac{d\tilde{E}}{2\pi}G^{<}(x_{1},x_{2},\boldsymbol{k'},\tilde{E})}\left(Pr\dfrac{1}{E-\tilde{E}-\hbar\omega_{LO}}-Pr\dfrac{1}{E-\tilde{E}+\hbar\omega_{LO}}\right)\right]
\end{gathered}
 \label{eq:popselfenergyret1D}
\end{equation}
\end{widetext}
for bulk systems. This expression has been used extensively in modeling quantum cascade lasers~\cite{kubis2009theory} and LEDs~\cite{steiger2009modelling}. 
The POP phonon scattering self-energies for UTB devices reads 
\begin{widetext}
\begin{equation}
\begin{gathered}
{\displaystyle \Sigma^{<}_{pop}\left(\boldsymbol{{x}_{1}},\boldsymbol{{x}_{2}},k,E\right)}=\dfrac{e^{2}}{(2\pi)^{3}}\left(\dfrac{1}{\epsilon_{\infty}}-\dfrac{1}{\epsilon_{s}}\right)\dfrac{\hbar\omega_{LO}}{2\epsilon_{0}}\\
\times{\displaystyle \int dk^{'}}I(k,k',\boldsymbol{{x}_{1}},\boldsymbol{{x}_{2}})\left[\vphantom{\dfrac{1}{2}}N_{ph}G^{<}\left(\boldsymbol{{x}_{1}},\boldsymbol{{x}_{2}},k',E-\hbar\omega_{LO}\right)\right.
\left.+\left(1+N_{ph}\right)G^{<}\left(\boldsymbol{{x}_{1}},\boldsymbol{{x}_{2}},k',E+\hbar\omega_{LO}\right)\vphantom{\dfrac{1}{2}}\right]
\end{gathered}
 \label{eq:popselfenergyless2D}
\end{equation}
\begin{equation} 
\begin{gathered}
{\displaystyle \Sigma^{R}_{pop}\left(\boldsymbol{{x}_{1}},\boldsymbol{{x}_{2}},k,E\right)}=\dfrac{e^{2}}{(2\pi)^{3}}\left(\dfrac{1}{\epsilon_{\infty}}-\dfrac{1}{\epsilon_{s}}\right)\dfrac{\hbar\omega_{LO}}{2\epsilon_{0}}
\times{\displaystyle \int dk^{'}}I(k,k',\boldsymbol{{x}_{1}},\boldsymbol{{x}_{2}})\left[\vphantom{\dfrac{1}{2}}(1+N_{ph})G^{R}\left(\boldsymbol{{x}_{1}},\boldsymbol{{x}_{2}},k',E-\hbar\omega_{LO}\right)\right.\\
+N_{ph}G^{R}\left(\boldsymbol{{x}_{1}},\boldsymbol{{x}_{2}},k',E+\hbar\omega_{LO}\right)+\dfrac{1}{2}G^{<}(\boldsymbol{{x}_{1}},\boldsymbol{{x}_{2}},k',E-\hbar\omega_{LO})
-\dfrac{1}{2}G^{<}(\boldsymbol{{x}_{1}},\boldsymbol{{x}_{2}},k',E+\hbar\omega_{LO})\\
+i{\displaystyle \int}\dfrac{d\tilde{E}}{2\pi}G^{<}\left(\boldsymbol{{x}_{1}},\boldsymbol{{x}_{2}},k',\tilde{E}\right)
\left.\left(Pr\dfrac{1}{E-\tilde{E}-\hbar\omega_{LO}}-Pr\dfrac{1}{E-\tilde{E}+\hbar\omega_{LO}}\right)\right]
\end{gathered}
\label{eq:popselfenergyret2D}
\end{equation}
\end{widetext}
where 
\begin{equation}
\begin{gathered}
I(k,k',\boldsymbol{x}_{1},\boldsymbol{x}_{2})=\\
\left\{ \begin{gathered}
\pi\left[\sqrt{\left(k-k'\right)^{2}+\zeta^{-2}}\left|\boldsymbol{{x}_{1}}-\boldsymbol{{x}_{2}}\right|+\dfrac{\left(k-k'\right)^{2}\left|\boldsymbol{{x}_{1}}-\boldsymbol{{x}_{2}}\right|}{\sqrt{\left(k-k'\right)^{2}+\zeta^{-2}}}\right]\\
\times K_{1}\left(\sqrt{\left(k-k'\right)^{2}+\zeta^{-2}}\left|\boldsymbol{{x}_{1}}-\boldsymbol{{x}_{2}}\right|\right),\;\left|\boldsymbol{{x}_{1}}-\boldsymbol{{x}_{2}}\right|\neq0\\
\pi\left[1+\dfrac{\left(k-k'\right)^{2}}{\left(k-k'\right)^{2}+\zeta^{-2}}\right],\;\left|\boldsymbol{{x}_{1}}-\boldsymbol{{x}_{2}}\right|=0
\end{gathered}\right.
\end{gathered}
\end{equation}
$K_1$ is the Bessel-K function of order 1. 
Finally, POP scattering self-energies in nanowires can be written as 
\begin{equation} 
\begin{gathered}
{\displaystyle \Sigma^{<}_{pop}\left(\vec{x}_{1},\vec{x}_{2},E\right)}=\dfrac{e^{2}}{(2\pi)^{3}}\left(\dfrac{1}{\epsilon_{\infty}}-\dfrac{1}{\epsilon_{s}}\right)\dfrac{\hbar\omega_{LO}}{2\epsilon_{0}}\\
\times I(\vec{x}_{1},\vec{x}_{2})\left[\vphantom{\dfrac{1}{2}}N_{ph}G^{<}\left(\vec{x}_{1},\vec{x}_{2},E-\hbar\omega_{LO}\right)\right.\\
+\left.\left(1+N_{ph}\right)G^{<}\left(\vec{x}_{1},\vec{x}_{2},E+\hbar\omega_{LO}\right)\vphantom{\dfrac{1}{2}}\right]
\end{gathered}
\label{eq:popselfenergyless3D}
\end{equation}
\begin{equation}
\begin{gathered}
{\displaystyle \Sigma^{R}_{pop}\left(\vec{x}_{1},\vec{x}_{2},E\right)}=\dfrac{e^{2}}{(2\pi)^{3}}\left(\dfrac{1}{\epsilon_{\infty}}-\dfrac{1}{\epsilon_{s}}\right)\dfrac{\hbar\omega_{LO}}{2\epsilon_{0}}I(\vec{x}_{1},\vec{x}_{2})\\
\times\left[\vphantom{\dfrac{1}{2}}(1+N_{ph})G^{R}\left(\vec{x}_{1},\vec{x}_{2},E-\hbar\omega_{LO}\right)\right.\\
\left.+N_{ph}G^{R}\left(\vec{x}_{1},\vec{x}_{2},E+\hbar\omega_{LO}\right)\right.\\
+\dfrac{1}{2}G^{<}(\vec{x}_{1},\vec{x}_{2},E-\hbar\omega_{LO})
-\dfrac{1}{2}G^{<}(\vec{x}_{1},\vec{x}_{2},E+\hbar\omega_{LO})\\
+i{\displaystyle \int}\dfrac{d\tilde{E}}{2\pi}G^{<}\left(\vec{x}_{1},\vec{x}_{2},\tilde{E}\right)\\
\left.\times\left(Pr\dfrac{1}{E-\tilde{E}-\hbar\omega_{LO}}-Pr\dfrac{1}{E-\tilde{E}+\hbar\omega_{LO}}\right)\right]
\end{gathered}
 \label{eq:popselfenergyret3D}
\end{equation}
where 
\begin{equation}
\begin{gathered}
I(\vec{x}_{1},\vec{x}_{2})=\\
\left\{ \begin{gathered}
\dfrac{4\pi^{2}}{a}\left[\dfrac{1}{2\left(\zeta^{2}\left(\dfrac{\pi}{a}\right)^{2}+1\right)}-\dfrac{3a}{2\zeta\pi}tan^{-1}\left(\dfrac{\zeta\pi}{a}\right)+1
\vphantom{\dfrac{1}{2\left(\zeta^{2}\left(\dfrac{\pi}{a}\right)^{2}+1\right)}}\right],\\
\forall\;\left|\vec{x}_{1}-\vec{x}_{2}\right|=0\\
\dfrac{\pi^{2}}{\zeta}\left(\dfrac{2\zeta}{\left|\vec{x}_{1}-\vec{x}_{2}\right|}-1\right)e^{-\left|\vec{x}_{1}-\vec{x}_{2}\right|/\zeta},\;\left|\vec{x}_{1}-\vec{x}_{2}\right|\neq0
\end{gathered}\right.
\end{gathered}
\end{equation}
Similar to the charged impurity scattering self-energies derivation, the phonon momentum integration is assumed to run over $(-\infty,\infty)$. 
In the numerical implementation of the retarded self-energies for inelastic scattering on polar optical phonons, the principal value integrals are neglected due to their minimal contribution but rather high numerical load.~\cite{esposito2009quantum}

\subsection{Band tails and band gap narrowing} \label{section:urbach}

The Urbach band tail parameters ($E_{Urbach}$) are extracted from the slope of the exponentially decaying density of states below the band edge
\begin{equation}
    E_{Urbach}=\dfrac{\left(E_{1}-E_{2}\right)}{log\dfrac{Im\left[G^{R}(\vec{x},\vec{x},E_{1})\right]}{Im\left[G^{R}(\vec{x},\vec{x},E_{2})\right]}}
 \label{eq:Urbachtail}
\end{equation}
where $E_1$ is one LO phonon energy below the ballistic band edge and $E_2$ is an integer number of phonon energies below $E_1$. 
The Green's functions are evaluated at a position $\vec{x}$ in the center of the device. 
To compensate for small numerical fluctuations of the band tail, Eq.~\ref{eq:Urbachtail} is solved for several $E_2$ ranging between 2 and 4 LO phonon energies below $E_1$. 
The average of these values is then used for the actual $E_{Urbach}$ result. 

The real part of the retarded scattering self-energy provides an energy shift and the imaginary part provides a energy broadening of the electronic states. 
Band gap narrowing is determined by running two sets of simulations - the first simulation solves for only the imaginary part of the retarded self-energy and its real part is set to zero. 
In that case, the band edges agree with those of ballistic calculations. 
In the second NEGF calculation the full retarded self-energy is solved. 
Note that the retarded self-energies only then fulfills the Kramers-Kr\"{o}nig relation. 
The band edges of this second case are defined as those energies where the density of states amplitude agrees with the band edge density of state amplitude of the first case, i.e. of the NEGF solution with purely imaginary retarded self-energy. 
The band gap narrowing is assumed to equal the differences in the band edges of those two cases. 
\begin{equation}
    E_{BGN}=E_{edge}\left(Re\left(\Sigma^{R}\right)=0\right)-E_{edge}\left(Re\left(\Sigma^{R}\right)\neq0\right)
\end{equation}

\subsection{Scattering rates from retarded self-energies}

Self-energies are verified by comparing their on-shell scattering rates against corresponding Fermi's golden rule results. 
On-shell scattering rates are computed by performing a basis transformation on the imaginary part of the retarded self-energy from the Wigner coordinate ($x-x'$) to the momentum space ($x$ being the transport direction). 
In many-band tight-binding, for a given energy-momentum (E-k) tuple, multiple on-shell target momentum values can be available to Fourier transform $x-x'$. 
All their contributions need to be summed up for each respective E-k tuple.
For ultra-thin bodies and nanowires, the self-energy is first transformed into the space spanned by the corresponding cross sectional modes $\zeta_{i}$. 
Then they are Fourier transformed to get both the inter and intra-mode scattering rates.

Following are the on-shell scattering rates as functions of retarded self-energies for bulk~\cite{wacker2002semiconductor}, ultra-thin bodies and nanowires
 \begin{equation}
 \begin{gathered}
     \Gamma\left(\boldsymbol{k},k_{x},E\right)=-\sum_{k_{x}}\left[\dfrac{2}{\hbar a}{\displaystyle \int_{-\infty}^{\infty}}d(x-x')e^{ik_{x}(x-x')}\right.\\
     \left.\times{\displaystyle \Sigma^{R}\left(\boldsymbol{k},x-x',\dfrac{x+x'}{2}\right)}\right]
 \end{gathered}
 \end{equation}
\begin{equation}
\begin{gathered}
\Gamma_{ij}\left(k,k_{x},E\right)=-\sum_{k_{x}}\left[\dfrac{2}{\hbar a}{\displaystyle \int_{-\infty}^{\infty}}d(x-x')e^{ik_{x}(x-x')}\right.\\
\left.{\displaystyle \tilde{\Sigma}^{R}_{ij}\left(k,x-x',\dfrac{x+x'}{2}\right)}\right]
\end{gathered}
\end{equation}
\begin{equation}
\begin{gathered}
\Gamma_{ij}\left(k_{x},E\right)=-\sum_{k_{x}}\left[\dfrac{2}{\hbar a}{\displaystyle \int_{-\infty}^{\infty}}d(x-x')e^{ik_{x}(x-x')}\right.\\
\left.{\displaystyle \tilde{\Sigma}^{R}_{ij}\left(x-x',\dfrac{x+x'}{2}\right)}\right]
\end{gathered}
\end{equation}
\begin{equation}
\tilde{\Sigma}^{R}_{ij}=V_i\Sigma^{R}V^{\dagger}_j
\end{equation}
where $\tilde{\Sigma}^{R}$ is the mode-space self-energy, $V$ is the eigenmode transformation matrix and $a$ is the lattice constant. 
Note that inelastic scattering in NEGF yields band tails in the band gap. 
Since those are not part of the linear response Fermi's golden rule in the next section, all NEGF-based scattering rate results are truncated for energies in the band gap.

\subsection{Scattering rates from Fermi's golden rule}
\label{Sec:FGR}
Due to the non-local nature of scattering, the Fermi's golden rule formulas are separately derived for each degree of confinement similar to Ref.~\onlinecite{goodnick1988effect}. 
Envelope wave functions of the form 
\begin{equation}
\psi(\vec{r})=\dfrac{e^{i\vec{k}\cdot\vec{r}}}{\sqrt{V}}
\end{equation}
for bulk, 
\begin{equation}
\psi(\vec{r})=\dfrac{e^{i\boldsymbol{k\cdot r}}}{\sqrt{A}}\zeta_{i}(z)
\end{equation}
for ultra-thin body, and
\begin{equation}
\psi(\vec{r})=\dfrac{e^{ikx}}{\sqrt{L}}\zeta_{i}(y,z)
\end{equation}
for nanowires are assumed. 
The scattering rate is solved with Fermi's golden rule for the bulk case with
\begin{equation}
\begin{gathered}
\dfrac{1}{\tau_{bulk}}=\dfrac{2\pi}{\hbar}\sum_{q}\left|\mel{\vec{k}\pm \vec{q}}{H}{\vec{k}}\right|^{2}\\
\times f_{initial}(1-f_{final})\delta(E_{final}-E_{initial})
\end{gathered}
\end{equation}
Here, $\vec{q}$  is the transferred momentum during scattering. 
$f_{initial}$ and $f_{final}$ are the electronic distribution functions for initial and final states. 
Scattering rates for ultra-thin body and nanowire are accordingly given with
\begin{equation}
\begin{gathered}
\dfrac{1}{\tau_{UTB,i,j}}=\dfrac{2\pi}{\hbar}\sum_{q}\left|\mel{\boldsymbol{k}\pm \boldsymbol{q},j}{H}{\boldsymbol{k},i}\right|^{2}\\
\times f_{initial}(1-f_{final})\delta(E_{final}-E_{initial})
\end{gathered}
\end{equation}
\begin{equation}
\begin{gathered}
\dfrac{1}{\tau_{wire,i,j}}=\dfrac{2\pi}{\hbar}\sum_{q}\left|\mel{k\pm q,j}{H}{k,i}\right|^{2}\\
\times f_{initial}(1-f_{final})\delta(E_{final}-E_{initial})
\end{gathered}
\end{equation}
where $i$ and $j$ are initial and final UTB and nanowire modes, respectively.
The transition element for bulk, ultra-thin bodies and nanowires read
    \begin{equation}
\left|\mel{\vec{k}\pm \vec{q}}{H}{\vec{k}}\right|^{2}=U_{\vec{q}}^{2}
\end{equation}
  \begin{equation}
  \begin{gathered}
  \left|\mel{\boldsymbol{k}\pm \boldsymbol{q},j}{H}{\boldsymbol{k},i}\right|^{2}=\\
  U_{\boldsymbol{q}}^{2}{\displaystyle \int_{-\infty}^{\infty}}dz{\displaystyle \int_{-\infty}^{\infty}}dz'\rho_{ij}(z)\rho_{ij}^{*}(z')e^{iq(z-z')}\\
  \end{gathered}
  \end{equation}
  \begin{equation}
      \rho_{ij}(z)=\zeta_{i}^{*}(z)\zeta_{j}(z)
  \end{equation}
and
\begin{equation}
\begin{gathered}
\left|\mel{k\pm q,j}{H}{k,i}\right|^{2}=\\
U_{q}^{2}{\displaystyle \int_{-\infty}^{\infty}}\boldsymbol{dr} \displaystyle{ \int_{-\infty}^{\infty}}\boldsymbol{dr'}\rho_{ij}(r)\rho_{ij}^{*}(r')e^{i\boldsymbol{q}\cdot(\boldsymbol{r}-\boldsymbol{r}')}\\
\end{gathered}
\end{equation}
\begin{equation}
    \rho_{ij}(\boldsymbol{r})=\zeta_{i}^{*}(y,z)\zeta_{j}(y,z)
\end{equation}
where $U_q$ is the scattering potential of Eqs.~\ref{Eq:Uq_imp} or \ref{Eq:Uq_pop}.

Inserting Eq.~\ref{Eq:Uq_imp}, the Fermi's golden rule for electrons scattering on charged impurities in  bulk results in  
    \begin{equation}
\dfrac{1}{\tau_{imp}(E)}=\dfrac{2e^{4}m^{*}N_{D}}{\pi\hbar^{3}\epsilon_{0}^{2}}\dfrac{\sqrt{2m^{*}E}/\hbar}{\zeta^{-2}\left(\zeta^{-2}+\dfrac{8m^{*}E}{\hbar^{2}}\right)}
\end{equation}
The Fermi's golden rule for electrons in ultra-thin body mode $i$ scattering on charged impurities into mode $j$ reads 
    \begin{equation}
\dfrac{1}{\tau_{imp,ij}(E)}=\dfrac{2e^{4}N_{D}m^{*}}{\hbar^{3}\epsilon_{0}^{2}(2\pi)^{3}}{\displaystyle \int_{0}^{2\pi}d\theta F(\mathbf{|\boldsymbol{k}-\boldsymbol{k'}|},\theta)}
\end{equation}
where 
\begin{equation}
\begin{gathered}
\mathbf{|\boldsymbol{k}-\boldsymbol{k'}|}=\left[\dfrac{2m^{*}}{\hbar^{2}}\left(2E-E_{i}-E_{j}\right)\right.\\
\left.-\dfrac{4m^{*}}{\hbar^{2}}\sqrt{(E-E_{i})(E-E_{j})}cos\theta\right]^{1/2}.
\end{gathered}
\end{equation}
The form factor $F$ is given by 
\begin{equation}
F(\boldsymbol{q})={\displaystyle \int_{0}^{L_{z}}}{\displaystyle \int_{0}^{L_{z}}}dzdz'\rho_{ij}(z)\rho_{ij}(z')I(\boldsymbol{q},z,z')
\end{equation}
where
\begin{equation}
I(\boldsymbol{q},z,z')=\dfrac{\pi e^{-\left|z-z'\right|\sqrt{\boldsymbol{q}^{2}+\zeta^{-2}}}}{2\left(\boldsymbol{q}^{2}+\zeta^{-2}\right)}\left[\dfrac{1}{\sqrt{\boldsymbol{q}^{2}+\zeta^{-2}}}+|z-z'|\right]
\end{equation}
The total scattering rate that mode $i$ faces is a sum of all possible mode transitions  
\begin{equation}
\Gamma_{i}(E)={\displaystyle \sum_{j}\Gamma_{ij}(E)={\displaystyle \sum_{j}\dfrac{1}{\tau_{ij}(E)}}}
\label{eq:multimodalscatt}
\end{equation} 
   The Fermi's golden rule for electrons in nanowire mode $i$ scattering on charged impurities into mode $j$ can be written as
    \begin{equation}
        \dfrac{1}{\tau_{ij}(E)}=\dfrac{e^{4}N_{D}\sqrt{2m^{*}}}{\hbar^{2}\epsilon_{0}^{2}(2\pi)^{3}}\left(\dfrac{F(k-k')+F(k+k')}{\sqrt{E-E_{i}}}\right)
    \end{equation}
    Form factor $F$ is given by
\begin{equation}
\begin{gathered}
\displaystyle F(q)=\\
\iint_A \iint_A \mathbf{\boldsymbol{dr}}\mathbf{\boldsymbol{dr'}}\rho_{ij}^{*}(\mathbf{\boldsymbol{r}})\rho_{ij}(\mathbf{\boldsymbol{r'}})I(q,\mathbf{r,r'})
\end{gathered}
\end{equation}
where the integration area $A = \left\{ (x,y)|\;0<x<L_x,\;0<y<L_y \right\}$ and 
\begin{equation}
I(q,\mathbf{r,r'}) =\left\{ \begin{gathered}
\dfrac{\left|\mathbf{r}-\mathbf{r'}\right|}{2\sqrt{q^{2}+\zeta^{-2}}}K_{1}\left(\sqrt{q^{2}+\zeta^{-2}}\left|\mathbf{r-r'}\right|\right),\\
\;\left|\mathbf{r-r'}\right|\neq0\\
\dfrac{1}{2\left(q^{2}+\zeta^{-2}\right)},\;\left|\mathbf{r-r'}\right|=0
\end{gathered}\right.
\end{equation}
The total scattering rate of mode $i$ is a sum of all intermode transitions likewise Eqn.~\ref{eq:multimodalscatt}.

The Fermi's golden rule for electrons scattering on polar optical phonons in bulk are expressed below with absorption and emission branches given separately. 

\vspace{11pt}
 \textbf{\underline{Bulk - Absorption process}}
\begin{equation}
\dfrac{1}{\tau_{ab}(\vec{k})}=\dfrac{e^{2}m^{*}\hbar\omega_{LO}N_{ph}}{4\pi\hbar^{3}\epsilon_0|\vec{k}|}\left(\dfrac{1}{\epsilon_{\infty}}-\dfrac{1}{\epsilon_{s}}\right){\displaystyle \int_{q_{-}}^{q+}dq}\dfrac{q^{3}}{\left(q^{2}+\zeta^{-2}\right)^{2}}
\end{equation}
where the integration limits of $q\;(q-,q+)$ are 
\begin{equation}
\begin{gathered}
\left(\dfrac{2m^{*}}{\hbar^{2}}\right)^{1/2}\left[\sqrt{E+\hbar\omega_{LO}}-\sqrt{E}\right]\leq q\\
\leq\left(\dfrac{2m^{*}}{\hbar^{2}}\right)^{1/2}\left[\sqrt{E+\hbar\omega_{LO}}+\sqrt{E}\right]
\end{gathered}
\end{equation}

\textbf{\underline{Bulk - Emission process}}
\begin{equation}
\begin{gathered}
\dfrac{1}{\tau_{em}(\vec{k})}=\theta\left(E-\hbar\omega_{LO}\right)\dfrac{e^{2}m^{*}\hbar\omega_{LO}\left(1+N_{ph}\right)}{4\pi\hbar^{3}\epsilon_0|\vec{k}|}\\
\times\left(\dfrac{1}{\epsilon_{\infty}}-\dfrac{1}{\epsilon_{s}}\right){\displaystyle \int_{q_{-}}^{q+}dq}\dfrac{q^{3}}{\left(q^{2}+\zeta^{-2}\right)^{2}}
\end{gathered}
\end{equation}
where the integration limits of $q\;(q-,q+)$ are 
\begin{equation}
\begin{gathered}
\left(\dfrac{2m^{*}}{\hbar^{2}}\right)^{1/2}\left[\sqrt{E}-\sqrt{E-\hbar\omega_{LO}}\right]\leq q\\
\leq\left(\dfrac{2m^{*}}{\hbar^{2}}\right)^{1/2}\left[\sqrt{E}+\sqrt{E-\hbar\omega_{LO}}\right]
\end{gathered}
\end{equation}
and $\theta$ represents the Heaviside step function.

Total scattering rate is the sum of emission and absorption processes and is given by
\begin{equation}
\dfrac{1}{\tau(\vec{k})}=\dfrac{1}{\tau_{em}(\vec{k})}+\dfrac{1}{\tau_{ab}(\vec{k})}
\end{equation}

  In a similar way, the absorption and emission contributions to the Fermi's golden rule for electrons in the ultra-thin body mode $i$ scattering on polar optical phonons into mode $j$ can be expressed as
  
    \vspace{11pt}
    \textbf{\underline{UTB - Absorption process}}
    \begin{equation}
    \begin{gathered}
\dfrac{1}{\tau_{ij,abs}(E)}=\dfrac{4\pi e^{2}m^{*}\hbar\omega_{LO}}{\hbar^{3}\epsilon_0}\left(\dfrac{1}{\epsilon_{\infty}}-\dfrac{1}{\epsilon_{s}}\right)N_{ph}\\
\times \int_{0}^{2\pi}d\theta F_{abs}(\mathbf{|\boldsymbol{k}-\boldsymbol{k'}|},\theta)
\end{gathered}
\end{equation}
where 
\begin{equation}
\begin{gathered}
\mathbf{|\boldsymbol{k}-\boldsymbol{k'}|}=\left[2\boldsymbol{k}^{2} + \dfrac{2\left(\hbar\omega_{LO} + (E_{i}-E_{j})\right)}{\hbar^{2}}\right.\\
\left. -2\boldsymbol{k}\left[\boldsymbol{k}^{2} + \dfrac{2\left(\hbar\omega_{LO} + (E_{i}-E_{j})\right)}{\hbar^{2}}\right]^{1/2}cos\theta\right]^{1/2}
\end{gathered}
\end{equation}

\textbf{\underline{UTB - Emission process}}
    \begin{equation}
    \begin{gathered}
\dfrac{1}{\tau_{ij,emi}(E)}=\dfrac{4\pi e^{2}m^{*}\hbar\omega_{LO}}{\hbar^{3}\epsilon_0}\left(\dfrac{1}{\epsilon_{\infty}}-\dfrac{1}{\epsilon_{s}}\right)(N_{ph}+1)\\
\times\int_{0}^{2\pi}d\theta F_{emi}(\mathbf{| \boldsymbol{k}- \boldsymbol{k'}|},\theta)
\end{gathered}
\end{equation}
where 
\begin{equation}
\begin{gathered}
\mathbf{| \boldsymbol{k}-\boldsymbol{k'}|}=\left[2\boldsymbol{k}^{2} - \dfrac{2\left(\hbar\omega_{LO} - (E_{i}-E_{j}) \right)}{\hbar^{2}}\right.\\
\left.-2k\left[\boldsymbol{k}^{2} - \dfrac{2\left(\hbar\omega_{LO} - (E_{i}-E_{j})\right)}{\hbar^{2}}\right]^{1/2}cos\theta\right]^{1/2}
\end{gathered}
\end{equation}
Form factor $F_{abs/emi}$ is given by 
\begin{equation}
F_{abs/emi}(\boldsymbol{q})={\displaystyle \int_{0}^{L_{z}}}{\displaystyle \int_{0}^{L_{z}}}dzdz^{'}\rho_{ij}(z)\rho_{ij}(z')I(\boldsymbol{q},z,z')
\end{equation}
where
\begin{equation}
\begin{gathered}
I(\boldsymbol{q},z,z')=\dfrac{e^{-\sqrt{\boldsymbol{q}^{2}+\zeta^{-2}}\left|z-z'\right|}}{\sqrt{\boldsymbol{q}^{2}+\zeta^{-2}}}\\
\times\left[1-\dfrac{\left|z-z'\right|\zeta^{-2}}{2\sqrt{\boldsymbol{q}^{2}+\zeta^{-2}}}-\dfrac{\zeta^{-2}}{2\left(\boldsymbol{q}^{2}+\zeta^{-2}\right)}\right]
\end{gathered}
\end{equation}

The total scattering of electrons in the mode $i$ is the sum of absorption and emission processes 
\begin{equation}
\Gamma_{i}(E)={\displaystyle \sum_{j}\Gamma_{ij}(E)={\displaystyle \sum_{j}\dfrac{1}{\tau_{ij,abs}(E)} + \dfrac{1}{\tau_{ij,emi}(E)}}}
\label{eq:multimodalscatt}
\end{equation} 

   The rates for absorption and emission of polar optical phonons of electrons in the nanowire mode $i$ scattering into mode $j$ read
    
    \vspace{11pt}
    \textbf{\underline{Nanowire - Absorption process}}
    \begin{equation}
    \begin{gathered}
\dfrac{1}{\tau_{ij,abs}(E)}=\dfrac{e^{2}\hbar\omega_{LO}N_{ph}}{\hbar^{2}\epsilon_0}\left(\dfrac{1}{\epsilon_{\infty}}-\dfrac{1}{\epsilon_{s}}\right).\dfrac{2}{(2\pi)^{2}}.\\
\times\sqrt{2m^{*}}\left(\dfrac{F(k-k')+F(k+k')}{\sqrt{E-E_{j}+\hbar\omega_{LO}}}\right)
\end{gathered}
\end{equation}
\begin{equation}
k=\sqrt{\dfrac{2m^{*}\left(E-E_{i}\right)}{\hbar^{2}}}\;\; k'=\sqrt{\dfrac{2m^{*}\left(E-E_{j}+\hbar\omega_{LO}\right)}{\hbar^{2}}}
\end{equation}
    
 \textbf{\underline{Nanowire - Emission process}}
    \begin{equation}
    \begin{gathered}
\dfrac{1}{\tau_{ij,emi}(E)}=\dfrac{e^{2}\hbar\omega_{LO}\left(N_{ph}+1\right)}{\hbar^{2}\epsilon_0}\left(\dfrac{1}{\epsilon_{\infty}}-\dfrac{1}{\epsilon_{s}}\right)\dfrac{2}{(2\pi)^{2}}\\
\times\sqrt{2m^{*}}\left(\dfrac{F(k-k')+F(k+k')}{\sqrt{E-E_{j}-\hbar\omega_{LO}}}\right)
\end{gathered}
\end{equation}
\begin{equation}
k=\sqrt{\dfrac{2m^{*}\left(E-E_{i}\right)}{\hbar^{2}}}\;\; k'=\sqrt{\dfrac{2m^{*}\left(E-E_{j}-\hbar\omega_{LO}\right)}{\hbar^{2}}}
\end{equation}

Form factor $F$ is given by
\begin{equation}
\begin{gathered}
\displaystyle F(q)=\\
\iint_A \iint_A \mathbf{dr}\mathbf{dr'}\rho_{ij}^{*}(\mathbf{r})\rho_{ij}(\mathbf{r'})I(q,\mathbf{r,r'})
\end{gathered}
\end{equation}
where $A = \left\{ (x,y) |\;0 < x < L_x,\;0 < y < L_y \right\}$
\begin{equation}
I(q,\mathbf{r,r'}) =\left\{ \begin{gathered}
\left(\sqrt{q^{2}+\zeta^{-2}}\left|\mathbf{r-r'}\right|+\dfrac{q^{2}\left|\mathbf{r-r'}\right|}{\sqrt{q^{2}+\zeta^{-2}}}\right)\\
\times\dfrac{K_{1}\left(\sqrt{q^{2}+\zeta^{-2}}\left|\mathbf{r-r'}\right|\right)}{2},\\
\;\left|\mathbf{r-r'}\right|\neq0\\
\left(\dfrac{1}{2}+\dfrac{q^{2}}{2\left(q^{2}+\zeta^{-2}\right)}\right),\;\left|\mathbf{r-r'}\right|=0
\end{gathered}\right.
\end{equation}
Similar to the scattering rate formulas above the total scattering rate of electrons in the nanowire mode $i$ is the sum of absorption and emission into all possible modes $j$.

\subsection{Compensation factor for non-local scattering self-energies}
It was shown in Sec.~\ref{Sec:selfenergies} that the electron scattering on polar optical phonons and charged impurities are non-local. 
Accounting for that full nonlocality in atomistic NEGF is numerically very expensive.
Neglect of nonlocal scattering underestimates scattering and orbital diagonal scattering for a given  position violates selection rules~\cite{aeberhard2019challenges}.
In this work, the numerical implementation of the self-energies described above are limited to atom-blockdiagonals only, i.e. nonlocal effects are included only within the range of a single atom. 
Interorbital transitions are only allowed within the same atom.
To correct this underestimation of scattering, a compensation factor is deduced from the Fermi's golden rule results in Sec.~\ref{Sec:FGR}. 
The compensation factor is defined as the ratio of the Fermi's golden rule form factors for the local approximation vs. the full non-local formula. 
The compensation factor for UTBs and nanowires can be represented as 
\begin{equation}
\begin{gathered}
S_{UTB}=\\
\dfrac{{\displaystyle \int_{0}^{2\pi}}{\displaystyle \int_{0}^{L_{z}}}{\displaystyle \int_{0}^{L_{z}}}d\theta dzdz^{'}\rho_{ij}(z)\rho_{ij}(z')I(\mathbf{\mathrm{|\boldsymbol{k}-\boldsymbol{k'}|}},z,z')}{{\displaystyle \int_{0}^{2\pi}}{\displaystyle \int_{0}^{L_{z}}}{\displaystyle \int_{0}^{L_{z}}}d\theta dzdz^{'}\rho_{ij}(z)\rho_{ij}(z')I_{Local}(\mathbf{\mathrm{|\boldsymbol{k}-\boldsymbol{k'}|}},z,z')}
\end{gathered}
\label{eq:S_UTB}
\end{equation}
and
\begin{equation}
\begin{gathered}
S_{wire}=\\
\dfrac{{\displaystyle \iint_A \iint_A \mathbf{dr}\mathbf{dr'}\rho_{ij}^{*}(\mathbf{r})\rho_{ij}(\mathbf{r'})I(q,\mathbf{r,r'})}}{{\displaystyle \iint_A \iint_A \mathbf{dr}\mathbf{dr'}\rho_{ij}^{*}(\mathbf{r})\rho_{ij}(\mathbf{r'})I_{Local}(q,\mathbf{r,r'})}}
\end{gathered}
\label{eq:S_wire}
\end{equation}
For electrons scattering on charged impurities, the UTB and nanowire cases read
\begin{equation}
I_{Local,imp}(\boldsymbol{q},z,z')=\left\{ \begin{gathered}
\dfrac{\pi}{\left(\boldsymbol{q}^{2}+\zeta^{-2}\right)^{3/2}}\;\left|z-z'\right|=0\\
0,\;\left|z-z'\right|\neq0
\end{gathered}\right.
\end{equation}
and 
\begin{equation}
I_{Local,imp}(q,\mathbf{r,r'})=\left\{ \begin{array}{c}
\left(\dfrac{1}{2\left(q^{2}+\zeta^{-2}\right)}\right),\;\left|\mathbf{r-r'}\right|=0\\
0,\;\left|\mathbf{r-r'}\right|\neq0
\end{array}\right.
\end{equation}
For electrons scattering on polar optical phonons, they read
\begin{equation}
I_{Local,pop}(\boldsymbol{q},z,z')=\left\{ \begin{gathered}
\dfrac{1}{\sqrt{\boldsymbol{q}^{2}+\zeta^{-2}}}\left[1-\dfrac{\zeta^{-2}}{2\left(\boldsymbol{q}^{2}+\zeta^{-2}\right)}\right],\\
\;\left|z-z'\right|=0\\
0,\;\left|z-z'\right|\neq0
\end{gathered}\right.
\end{equation}
and
\begin{equation}
I_{Local,pop}(q,\mathbf{r,r'})=\left\{ \begin{array}{c}
\left(\dfrac{1}{2}+\dfrac{q^{2}}{2\left(q^{2}+\zeta^{-2}\right)}\right),\;\left|\mathbf{r-r'}\right|=0\\
0,\;\left|\mathbf{r-r'}\right|\neq0
\end{array}\right.
\end{equation}
The discretization of nonlocal scattering depends on the real space mesh size.
Self-energy matrices that are diagonal in the real space representation cover nonlocal scattering only within the volume represented by each single mesh point, i.e. each single atom.
Therefore, the compensation factors depend on the mesh spacing and vary with system dimensions. 
To accurately represent the zincblende lattices of this work, the mesh spacing is chosen to agree with the spacing between subsequent atomic planes in [100] direction (i.e. $(a_o/4)$). 
Accordingly, the integrals for the denominators of Eqs.~(\ref{eq:S_UTB}) and (\ref{eq:S_wire}) run over the atomic volume. 
Note that Fermi's golden rule formulations in bulk systems do not contain real space information - in contrast to the NEGF self-energies.
Thus, a formulation of Eqs.~(\ref{eq:S_UTB}) and (\ref{eq:S_wire}) for bulk is not possible.
Nevertheless, 50~nm UTBs with 20 electronic modes can mimic bulk behavior sufficiently well. 
The compensation factors of 50~nm UTB cases are therefore used in this work as bulk compensation factors.

\begin{figure}
\includegraphics[scale=0.31]{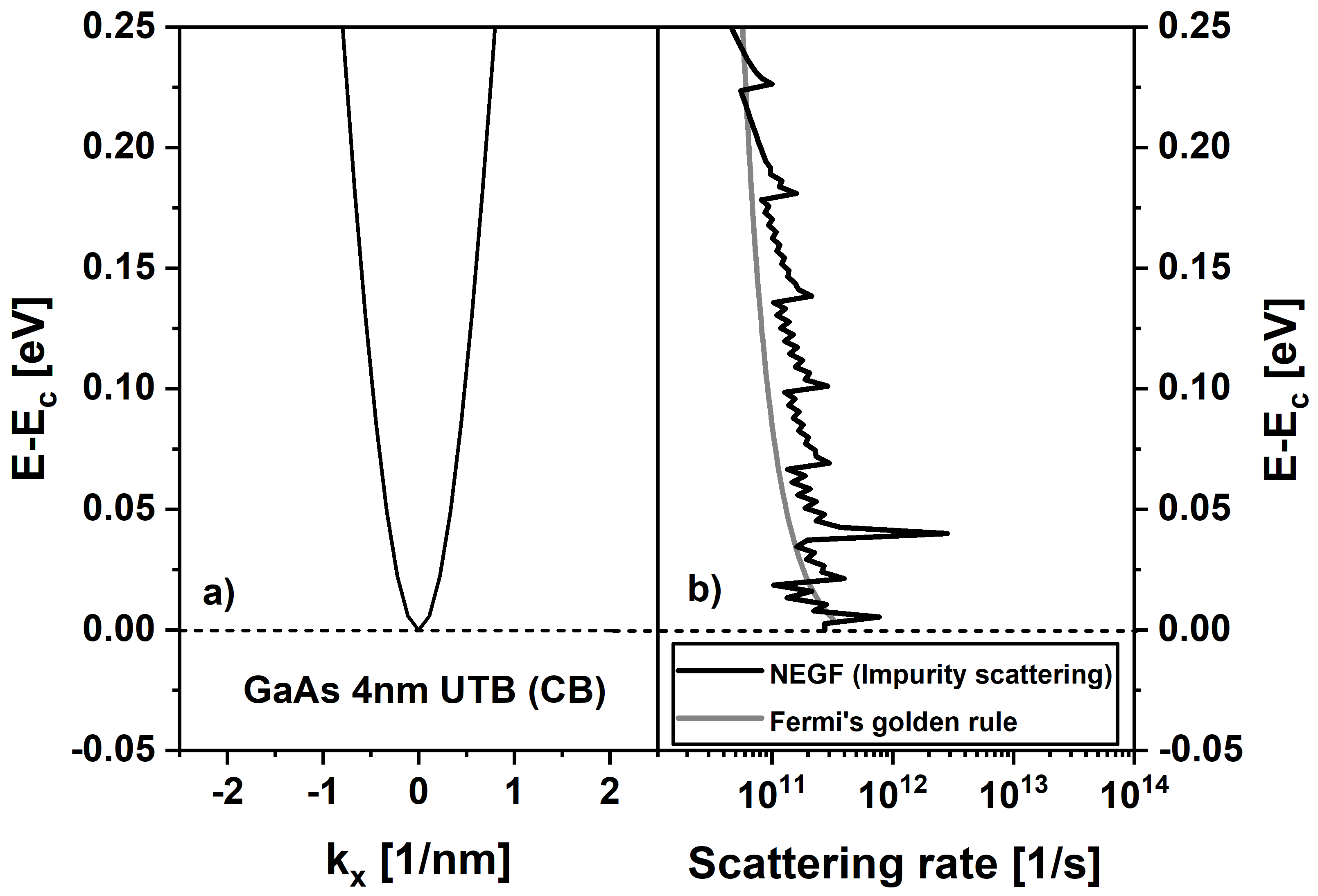}
\centering
\caption{a) Electronic dispersion of conduction band electrons of a 4~nm layer of GaAs in the [100] direction and in 10-band sp3d5s* tight binding representation. 
b) Scattering rates solved with NEGF (black) and with Fermi's golden rule (gray) for  conduction band electrons of (a) scattering on randomly distributed charged impurities of a concentration of $2\times10^{18} cm^{-3}$. The screening length is set to 3~nm.
The zero in energy is set to the bottom of the conduction band (dotted).
The spikes in the NEGF rates are due to limited resolution of the energy and momentum space and the Fourier transformation.}
\label{fig:rateimputb}
\end{figure}

\section{Results and discussions}

\subsection{Scattering rate comparison with Fermi's golden rule}

To verify the approximate treatment of nonlocal scattering self-energies of Sec.~\ref{Sec:selfenergies}, we benchmark the scattering rates of NEGF calculations against Fermi's golden rule in bulk, UTB and nanowire equilibrium systems.
The energy resolved scattering rate of electrons scattering on charged impurities follows the electronic density of states multiplied with the impurity scattering potential.
The latter introduces a $q^{-2}$ dependence of the rate, as can be seen in Fig.~\ref{fig:rateimputb} for conduction band electrons in a 4nm GaAs ultra-thin body.
The NEGF predicted scattering rate shows a good agreement with Fermi's golden rule. 
Note that the spikes in the NEGF rate result from the finite numerical resolution of the transverse momentum space.
Finer momentum meshes produce smoother NEGF scattering rates, but require significantly larger computational resources \cite{datta2005quantum}.

   \begin{figure}
\includegraphics[scale=0.31]{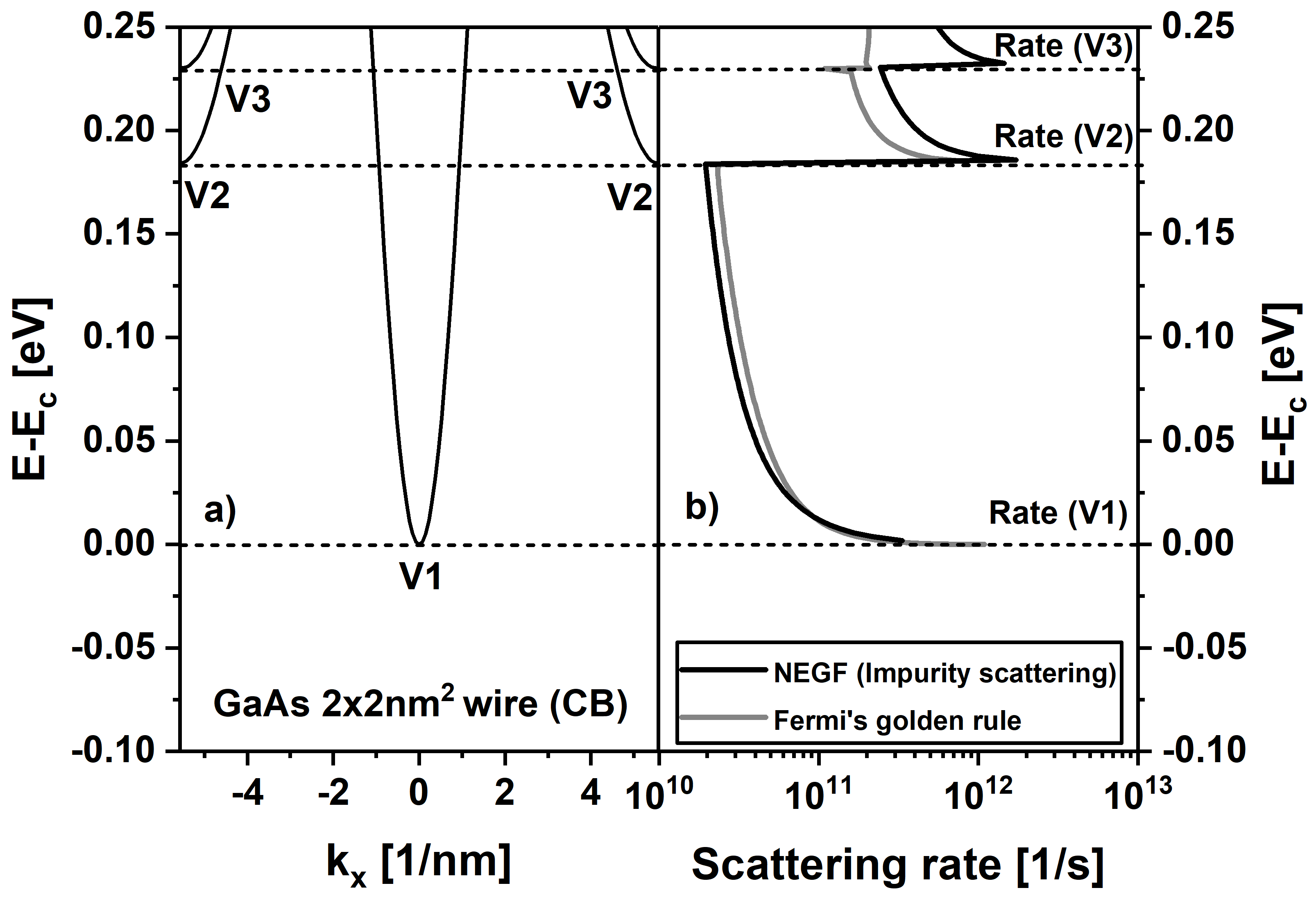}
\centering
\caption{a) Electronic dispersion of conduction band electrons of a $2\times2nm^{2}$ GaAs nanowire in 10-band sp3d5s* tight binding representation. 
Valleys at the $\Gamma$ point and the Brillouin zone boundary are labelled with V1, V2 and V3 and marked by dotted lines. 
The zero in energy is set to the bottom of the conduction band. b) Scattering rates solved with NEGF (black) and with Fermi's golden rule (gray) for conduction band electrons of a) scattering on randomly distributed charged impurities of a concentration of $2\times10^{18} cm^{-3}$. 
The screening length is set to $3~nm$. Fermi's golden rule shows good agreement with NEGF over a wide energy range.}
\label{fig:rateimpwireelectrons}
\end{figure}

Similarly, NEGF results of nanowire electrons scattering on charge impurities show good agreement with Fermi's golden rule (see Fig.~\ref{fig:rateimpwireelectrons}).
Short of a momentum degree of freedom and the related resolution challenges, the scattering rates smoothly follow the 1D density of states.
The steps in the scattering rates coincide with valley energies and mark the onset of additional inter-valley and intra-valley scattering.

\begin{figure}
\includegraphics[scale=0.31]{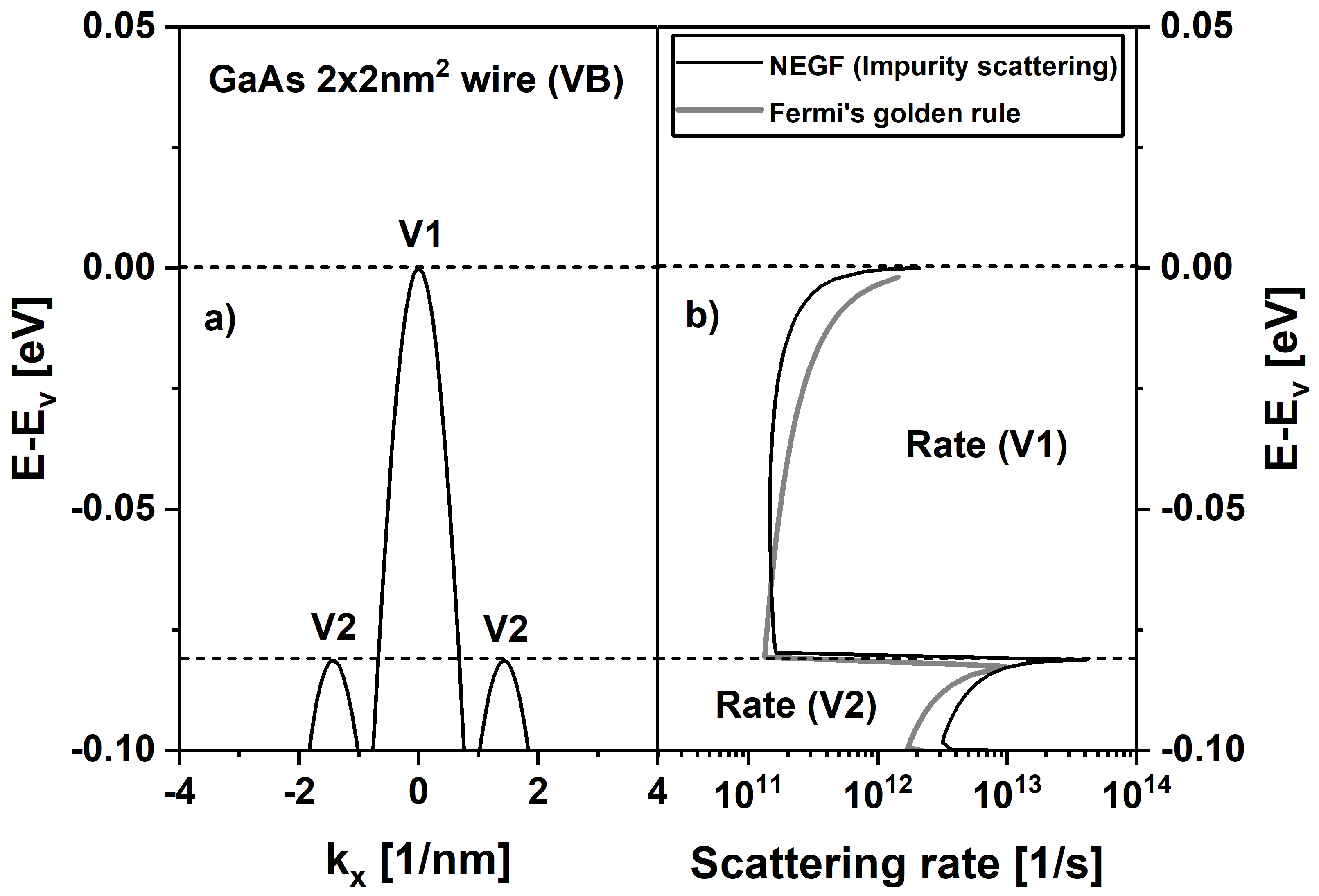}
\centering
\caption{a) Electronic dispersion of valence band electrons of a $2\times2nm^{2}$ GaAs nanowire in 10-band sp3d5s* tight binding representation. 
Valleys are labelled with V1 and V2 and marked by dotted lines. The zero in energy is set to the top of the valence band.
b) Scattering rates solved with NEGF (black) and with Fermi's golden rule (gray) for electrons of a) scattering on randomly distributed charged impurities of a concentration of $2\times10^{18} cm^{-3}$ The screening length is set to $3~nm$. Fermi's golden rule shows good agreement with NEGF over a wide energy range. 
Scattering rate steps show the onset of additional intra-valley and inter-valley scattering of valley V2.}
\label{fig:rateimpwireholes}
\end{figure}

GaAs valence band states are mainly composed of p-orbitals, whereas electronic wavefunctions at the conduction band edge are mainly s-orbital type \cite{tan2015tight}. 
Therefore, inter-orbital scattering is expected to be more important in the valence band. 
Figure~\ref{fig:rateimpwireholes} compares the scattering rates of GaAs valence band electrons. 
The good agreement between the NEGF results and Fermi's golden rule suggests the approximation of equal inter-orbital self-energy elements is appropriate.
It is worth to mention that neglecting inter-orbital scattering on the same atom typically reduces the scattering rates by about $3\times$.
Remaining deviations of NEGF and Fermi's golden rule results can be addressed to the effective mass dispersion assumed for the Fermi's golden rule results in contrast to the multi-band atomistic treatment in NEGF.
Note that the scattering rate at the valence band edge is smaller than the rate at the onset of the next valley (labeled V1 and V2 in Fig.~\ref{fig:rateimpwireholes}, respectively) due to the ratio of the valley effective masses and their impact on scattering density of states. 

    \begin{figure}
\includegraphics[scale=0.31]{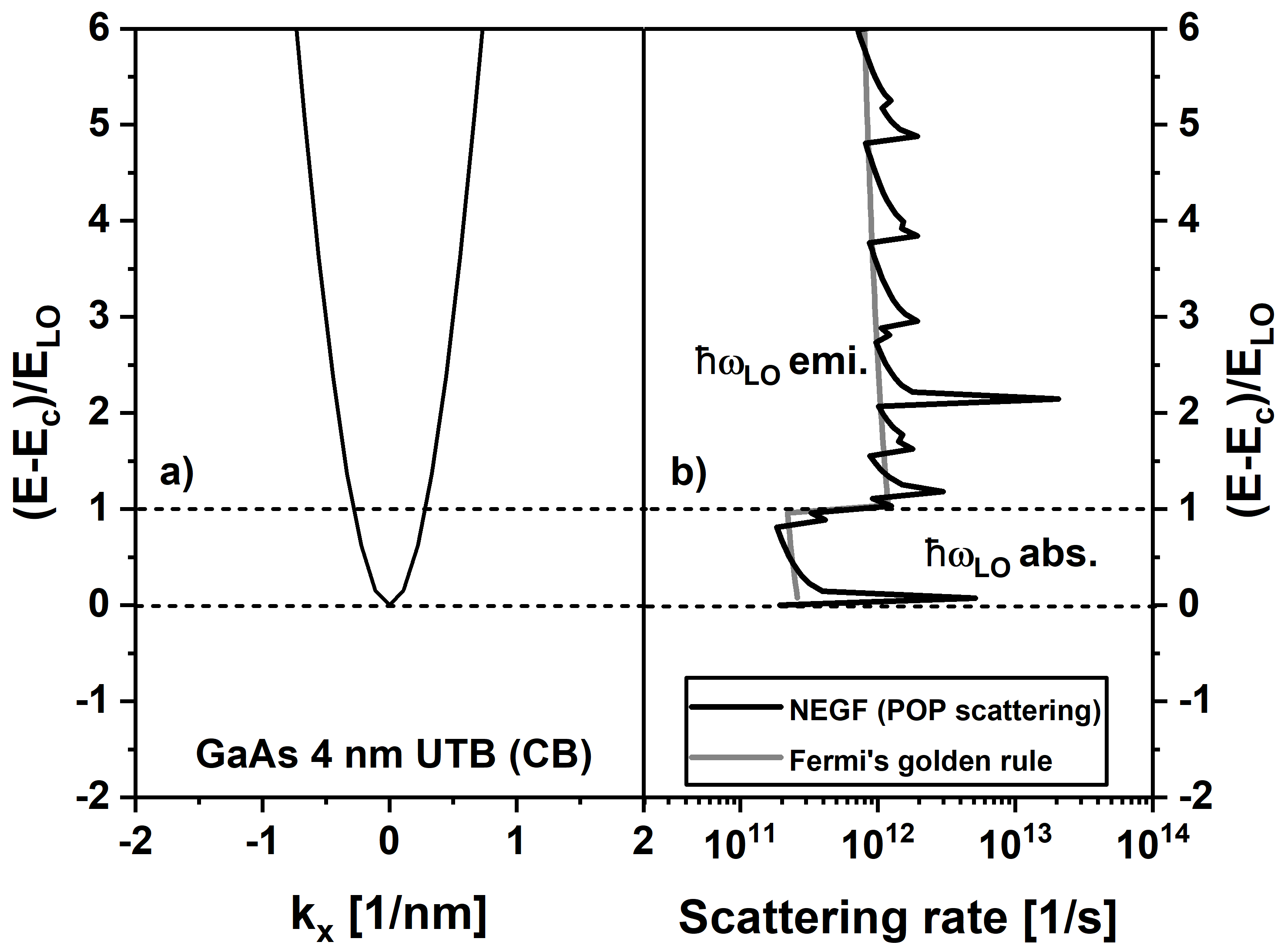}
\centering
\caption{a) Electronic dispersion of conduction band electrons of a $4~nm$ layer of GaAs in the [100] direction and in 10-band sp3d5s* tight binding representation. 
b) Scattering rates solved with NEGF (black) and with Fermi's golden rule (gray) for  conduction band electrons of (a) scattering on polar optical phonons. The screening length is set to 3~nm.
The zero in energy is set to the bottom of the conduction band (dotted).
Onset of phonon emission (labeled with "emi.") is observed at 1 LO phonon energy ($\hbar\omega_{LO}$) above conduction bandedge (dotted).
The spikes in the NEGF rates are due to limited resolution of the energy and momentum space and the Fourier transformation.}
\label{fig:ratepoputb}
\end{figure}

\begin{figure}
\includegraphics[scale=0.31]{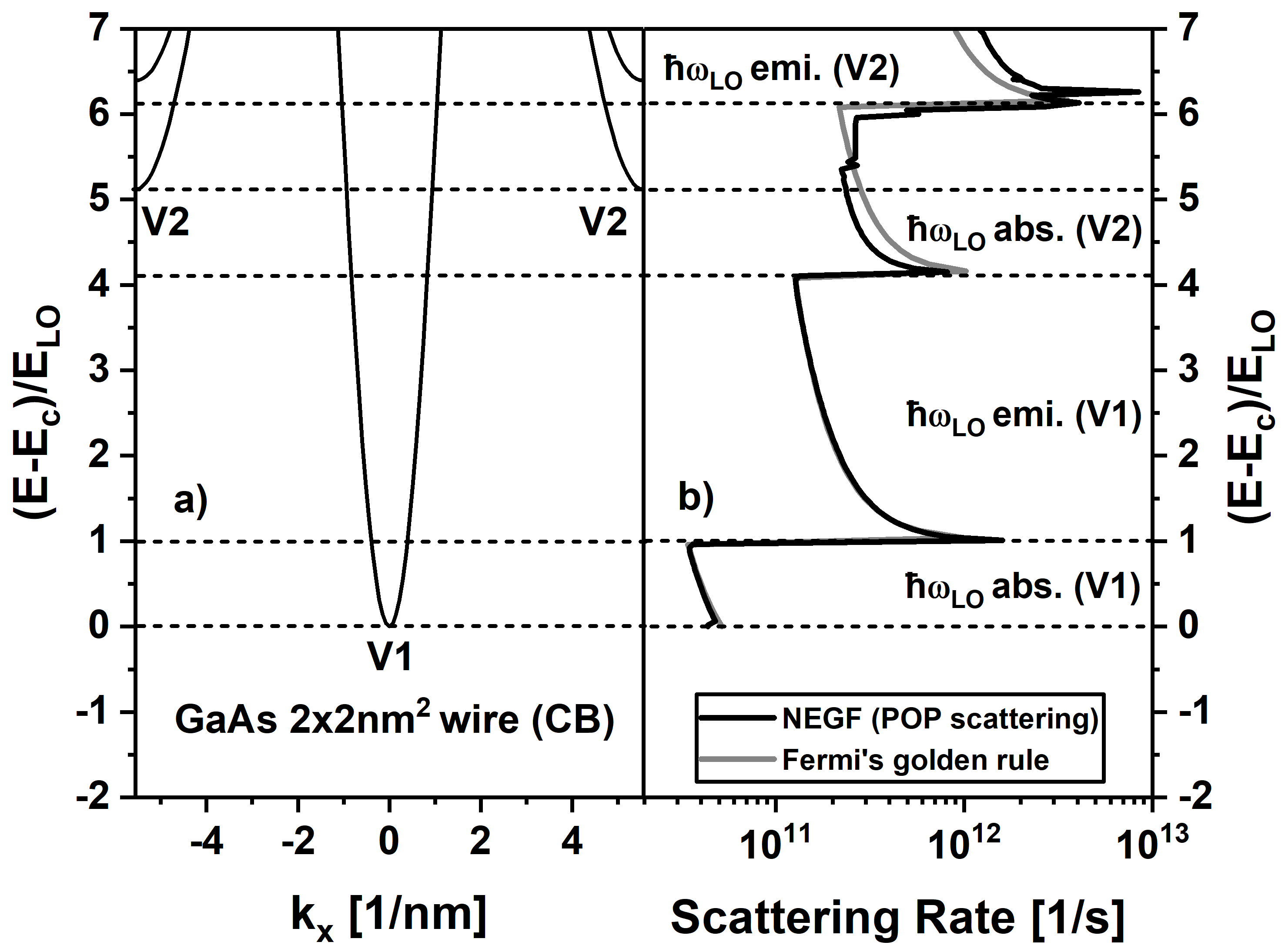}
\centering
\caption{a) Electronic dispersion of conduction band electrons of a $2\times2nm^{2}$ GaAs nanowire in 10-band sp3d5s* tight binding representation. 
Valleys at the $\Gamma$ point and the Brillouin zone boundary are labelled with V1 and V2 and marked by dotted lines. 
Marked with dotted lines are also energies that are 1 LO phonon energy below and above the respective valley bottom. 
The zero in energy is set to the bottom of the conduction band. 
b) Scattering rates solved with NEGF (black) and with Fermi's golden rule (gray) for conduction band electrons of a) scattering on LO phonons. 
The screening length is set to $3~nm$. Fermi's golden rule shows good agreement with NEGF over a wide energy range.
The onset of absorption (labeled with "abs.") and emission (labeled with "emi.") processes can be clearly observed for the conduction band valleys V1 and V2.}
\label{fig:ratepopwireelectrons}
\end{figure}

\begin{figure}
\includegraphics[scale=0.31]{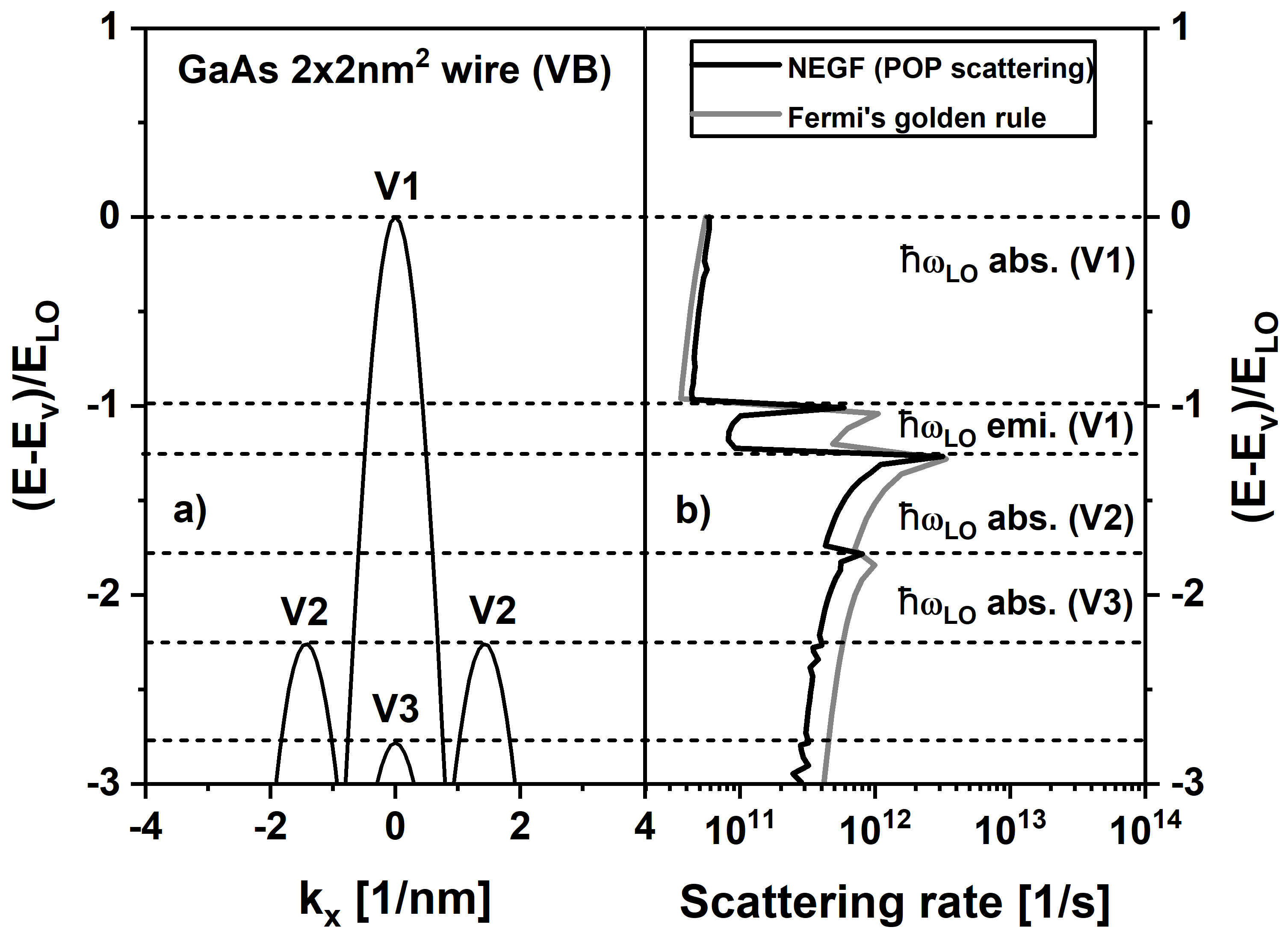}
\centering
\caption{a) Electronic dispersion of valence band electrons of a $2\times2nm^{2}$ GaAs nanowire in 10-band sp3d5s* tight binding representation. 
Valleys are labelled with V1, V2 and V3 and marked by dotted lines. Other dotted lines mark energies that are one LO phonon energy above or below the respective valley energy. The zero in energy is set to the top of the valence band.
b) Scattering rates solved with NEGF (black) and with Fermi's golden rule (gray) for electrons of a) scattering on LO phonons. The screening length is set to $3~nm$. Fermi's golden rule shows good agreement with NEGF over a wide energy range. 
Scattering rate steps mark the onset of absorption (labeled with "abs.") and emission (labeled with "emi.") processes  for the valleys V1, V2 and V3.}
\label{fig:ratepopwireholes}
\end{figure}

The rates for electron scattering on polar optical phonons predicted with NEGF and Fermi's golden rule of GaAs UTB conduction band and nanowire conduction and valence bands are shown in Figs.~\ref{fig:ratepoputb}, \ref{fig:ratepopwireelectrons}, and \ref{fig:ratepopwireholes}.
NEGF scattering rates in UTBs (Fig.~\ref{fig:ratepoputb}) exhibit spikes due to the numerical resolution of transverse momentum space similar to the impurity scattering in Fig.~\ref{fig:rateimputb}. 
All the results show steps when phonon emission and absorption processes at energy $\hbar\omega_{LO}$ above or below the various band and valley edges step in. 

The scattering rates in Fig.~\ref{fig:ratepopwireelectrons} clearly shows absorption and emission processses for valleys V1 and V2. 
For energies near the conduction band edge, the rate includes phonon absorption and emission of electrons in valley V1 only. 
Electrons with energies of $~4\hbar\omega_{LO}$ above the V1 edge can absorb LO phonons and scatter to valley V2 which results in an abrupt rate increase. 
Electrons in V2 with a total energy exceeding the V2 edge by one $\hbar\omega_{LO}$ can emit LO phonons within the same valley. 
Thus, the scattering self-energies capture inter-valley and intra-valley scattering processes.
Similar feature can be seen in Fig.~\ref{fig:ratepopwireholes} for electrons in the valence band.

\subsection{Urbach tail predictions vs. temperature, doping and confinement}
\begin{figure}
\includegraphics[scale=0.32]{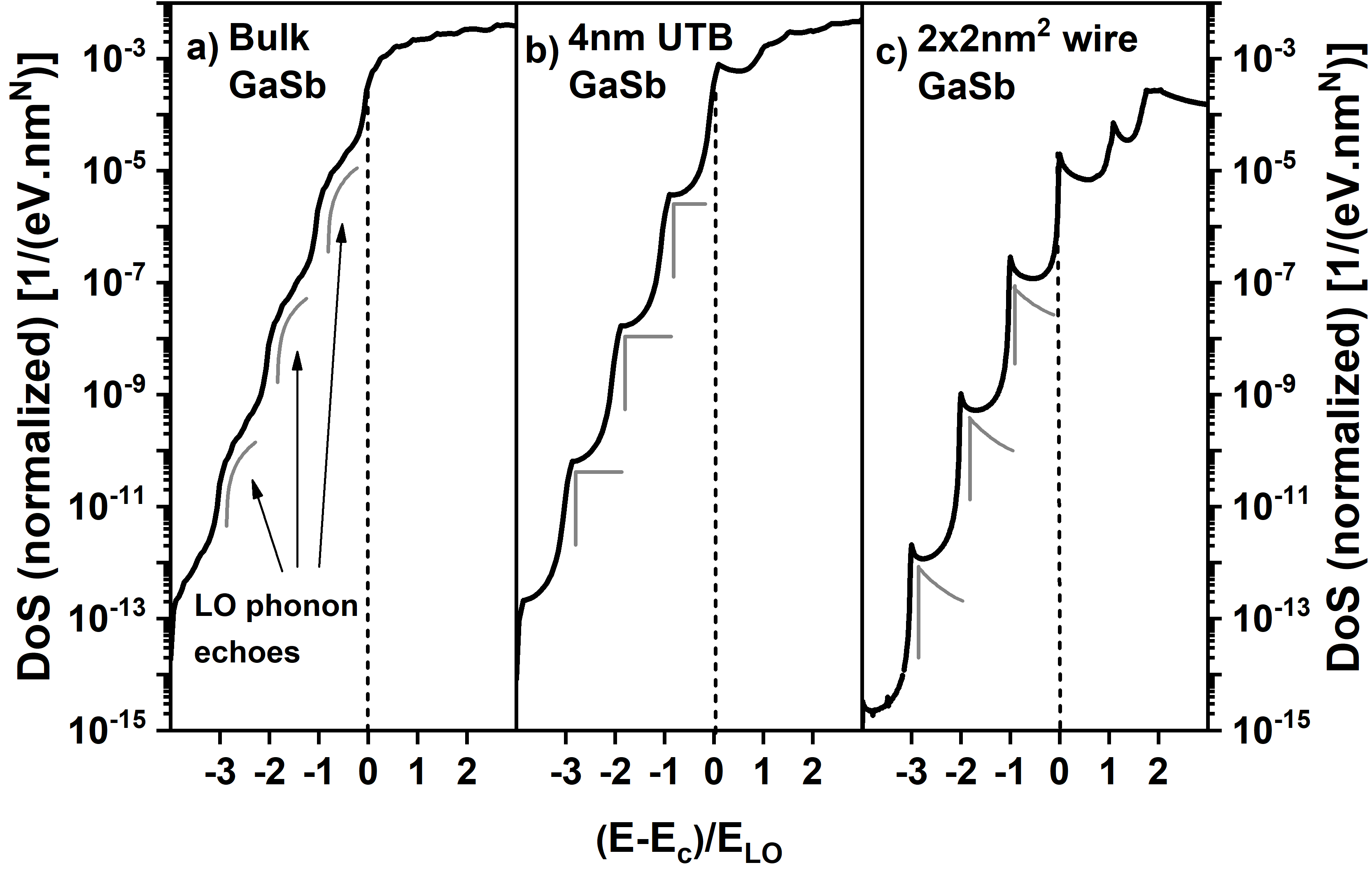}
\centering
\caption{Electronic density of states close to the conduction band edge of bulk GaSb, of a 4nm thick GaSb ultra-thin body and of a $2\times2nm^2$ GaSb nanowire in the presence of polar optical phonon and charged impurity scattering. 
The electronic density of states shows exponentially decaying band tails with LO phonon echoes of the density of states at the bandedge (indicated in gray). Therefore, these echos resemble the system's dimensionality indicated with the gray DOS sketches for bulk (a), UTB (b) and a nanowire (c), respectively. 
To ease the comparison, the real part of the retarded scattering self-energies is neglected here.}
\label{fig:bandtaildos}
\end{figure}

Urbach tails are an important aspect for optical experiments and the linewidth of absorption spectra \cite{dow1972toward, agarwal2014band, khayer2011effects}. 
Figure~\ref{fig:bandtaildos} shows the density of states of electrons in GaSb bulk, in a GaSb ultra-thin body and in a GaSb nanowire when scattering on charged impurities and polar optical phonons. 
The electronic density of states in the presence of scattering on polar optical phonons and charged impurities differs from the ballistic case that drops sharply at the ballistic bandedge (marked with dotted line in Fig.~\ref{fig:bandtaildos}). 
The considered scattering processes enhance the density of states below the bandedge into an exponential decaying band tail.
The shape of band tails is determined by the nature of the density of states close to the bandedge and the formation of LO phonon echos~\cite{antonioli1981intrinsic}. 

\begin{figure}
\includegraphics[scale=0.33]{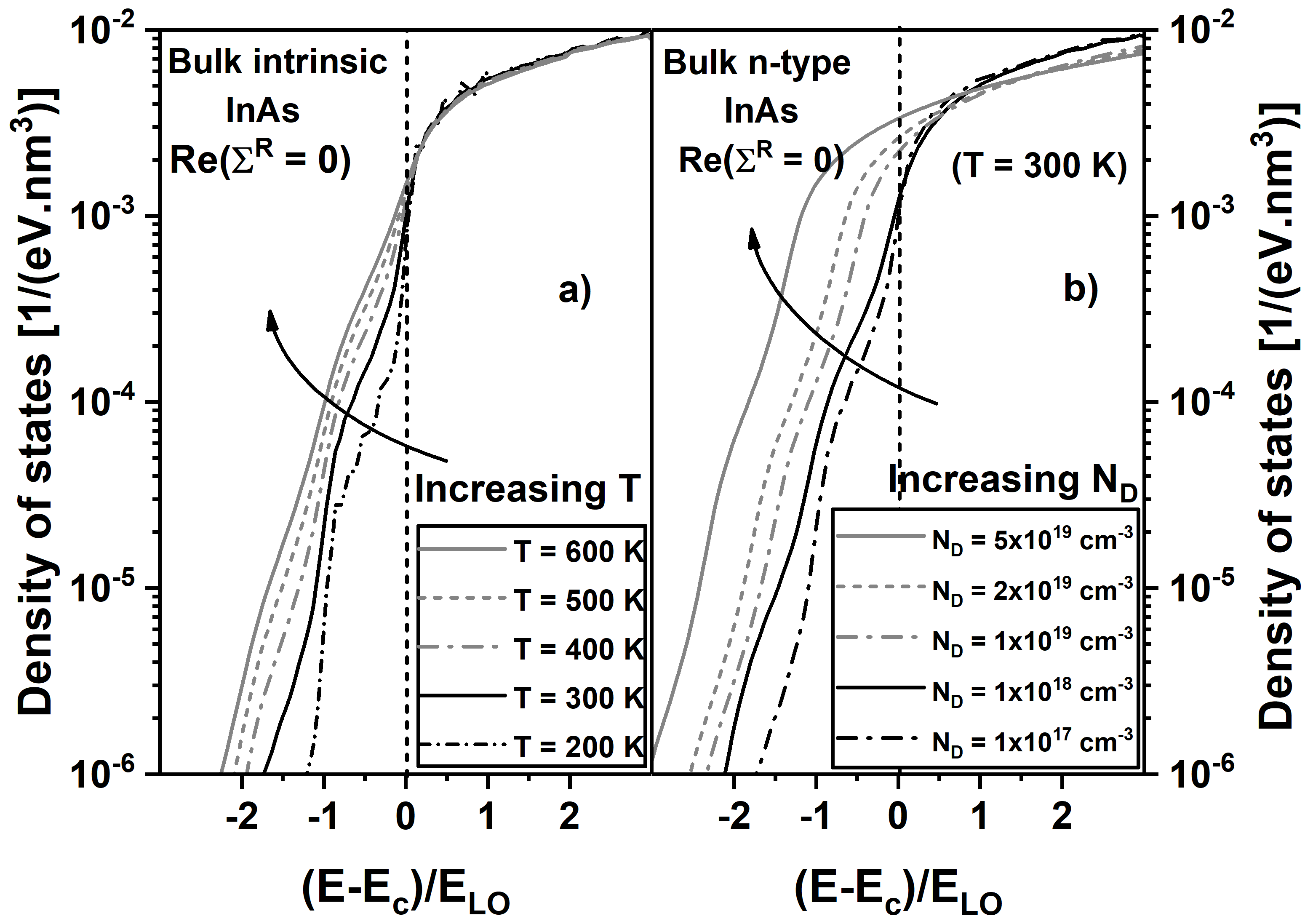}
\centering
\caption{a) Electronic density of states close to the conduction band edge of bulk InAs as a function of energy (normalized to LO phonon energy) for different temperatures. 
b) Same as (a), but at room temperature and with varying doping concentrations. 
Increasing the temperature and the doping concentration enhances the scattering-supported band tail formation. 
Elastic scattering that blurs the phonon echos is increased in this way as well.
To ease the comparison, the real part of the retarded scattering self-energies is neglected here.
}
\label{fig:dos_versus_doping_T_InAs}
\end{figure}

Figure~\ref{fig:dos_versus_doping_T_InAs} shows the variation of the conduction band tail with doping concentration and temperature. 
The phonon self-energies increase exponentially with temperature, which yields larger band tails (see Fig.~\ref{fig:dos_versus_doping_T_InAs}).
Impurity scattering is an elastic process and does not directly contribute to the band tail formation. 
However, once inelastic scattering on phonons creates a finite band tail, elastic scattering on charged impurities enhances the density of states at every energy in the band tail. 
Therefore, with increasing doping concentration, the band tail becomes larger and results in increasing Urbach parameters with doping. 

With increasing doping and increasing temperature, phonon echoes are gradually washed out.
This is due to increasing momentum randomization during scattering on impurities and polar optical phonons.
Increasing temperatures enlarge the Lindhard screening length $\zeta$ (see Eq.~\ref{eq:Lindhard}), which in turn supports larger differences of initial and final momentum in the exponents of the impurity and pop self-energies (see Eqs.~(\ref{eq:implessselfenergy1D}), (\ref{eq:implessselfenergy2D}), (\ref{eq:popselfenergyless1D}) and (\ref{eq:popselfenergyless2D})).
In this way the electronic dispersion at the band edge and with it the LO phonon echos gets blurred with increasing temperatures.

\begin{figure}
\includegraphics[scale=0.35]{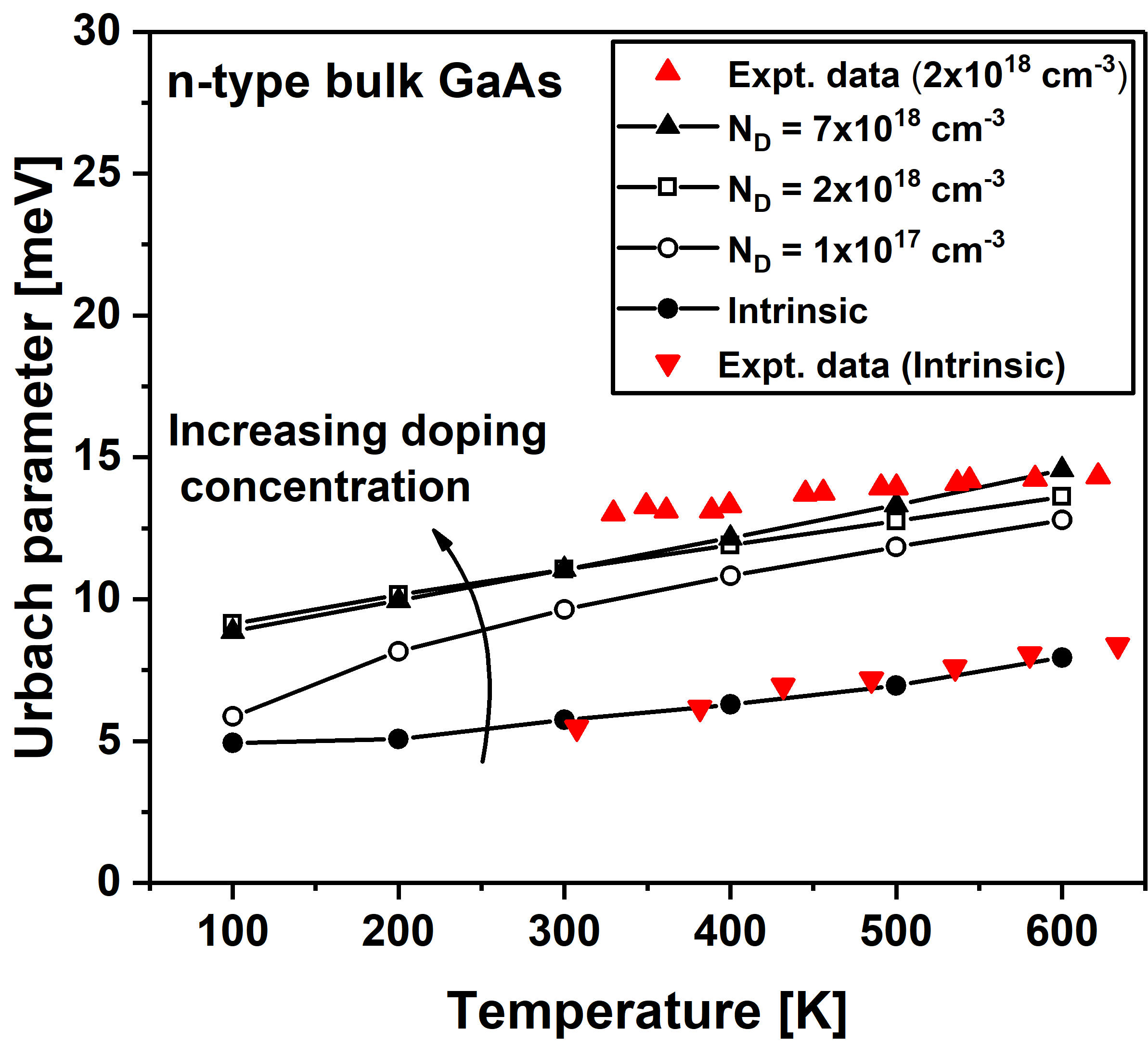}
\centering
\caption{Urbach parameter of bulk GaAs as a function of temperature for different n-type doping concentrations. 
The NEGF predictions show good agreement with the experimental data of Ref.~\onlinecite{johnson1995temperature}. 
Lines are meant to guide the eye.}
\label{fig:urbach_versus_expt_GaAs}
\end{figure}

As detailed in section~\ref{section:urbach}, the Urbach parameter is a scalar parameter that characterizes band tails.
Figure~\ref{fig:urbach_versus_expt_GaAs} shows the variation of the Urbach parameter in bulk GaAs as a function of temperature for different doping concentrations predicted with NEGF and compared against experimental data of Ref.~\onlinecite{johnson1995temperature}. 
The NEGF results agree quantitatively with the experimental data for the intrinsic material over a large temperature range.
The results agree well for the $2\times10^{18} cm^{-3}$ n-doped case.
The deviations in the doped case likely originate from scattering on disorder effects and neutral impurity potentials that come along with doping, but are neglected in the NEGF calculations. 

\begin{figure}
\includegraphics[scale=0.32]{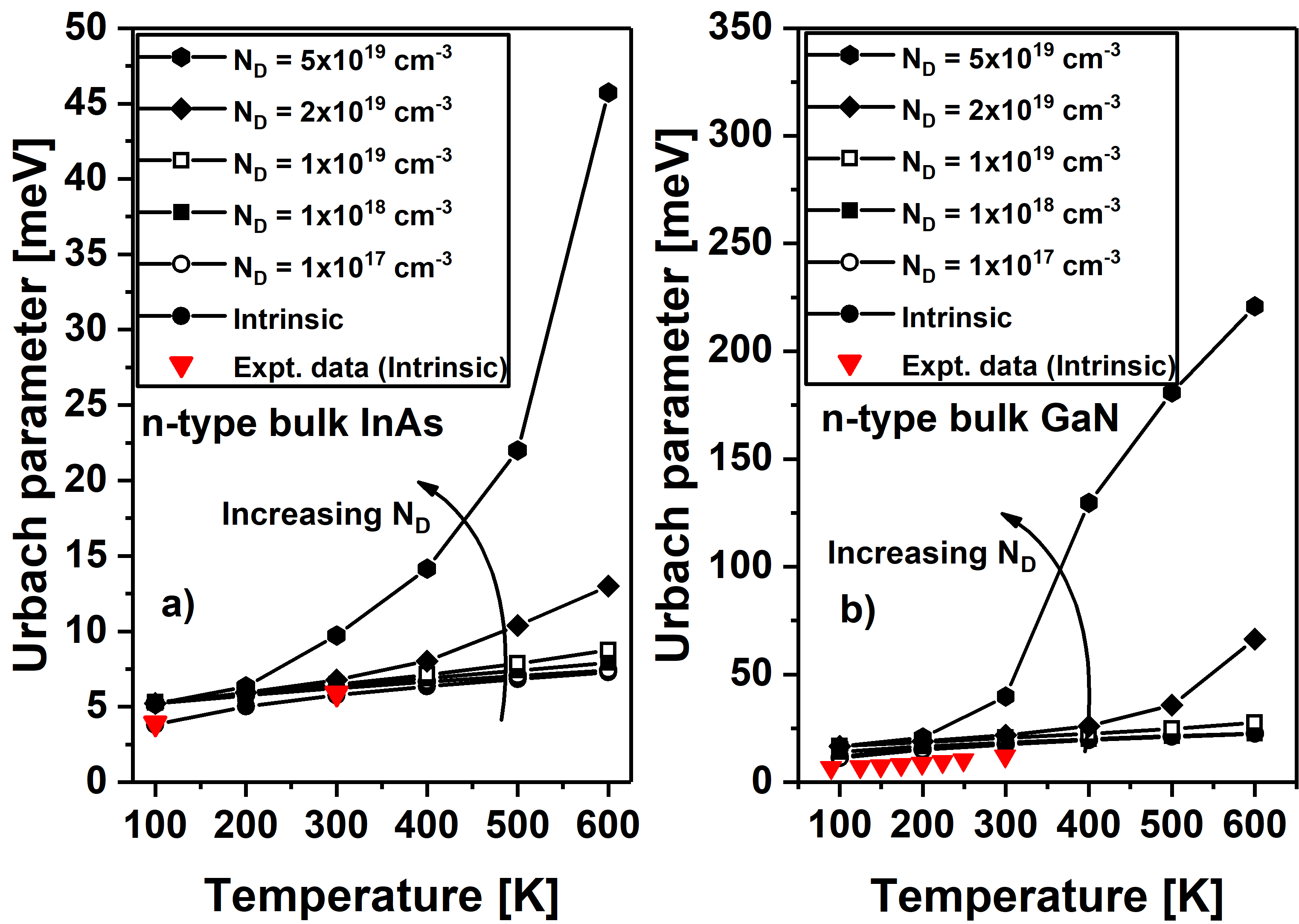}
\centering
\caption{Urbach parameter as a function of temperature for n-type bulk InAs a) and GaN b) for different doping concentrations. 
Simulation results for the intrinsic materials compare well with experimental data of Ref.~\onlinecite{antonioli1981intrinsic} for InAs and Ref.~\onlinecite{chichibu1997urbach} for GaN. 
With increasing doping concentration, the Urbach parameter increases more rapidly with temperature due to the combined effect of higher phonon and impurity scattering.
Lines are meant to guide the eye.}
\label{fig:urbach_versus_expt_GaN_InAs}
\end{figure}

Figures~\ref{fig:urbach_versus_expt_GaN_InAs} a) and b) show the Urbach parameter for InAs and GaN.
Also here, the NEGF predictions agree well with the experimental data of Refs.~\onlinecite{antonioli1981intrinsic} and \onlinecite{chichibu1997urbach}, for InAs and GaN, respectively.
The Urbach parameter value strongly depends on the strength of scattering potentials and available density of states near the band edge. 
The dielectric constants of GaN and the larger LO phonon energy of GaN ($92$ meV) supports stronger LO phonon scattering in GaN than in InAs (phonon energy of $30$ meV).
In addition, GaN has a conduction band effective mass ($0.2m^*$) that is larger than the one of InAs ($0.025m^*$) which enhances the density of states near its band edge. 
Consequently, Figs.~\ref{fig:urbach_versus_expt_GaN_InAs} show larger Urbach parameters with a stronger temperature dependence in GaN.

\begin{figure}
\includegraphics[scale=0.35]{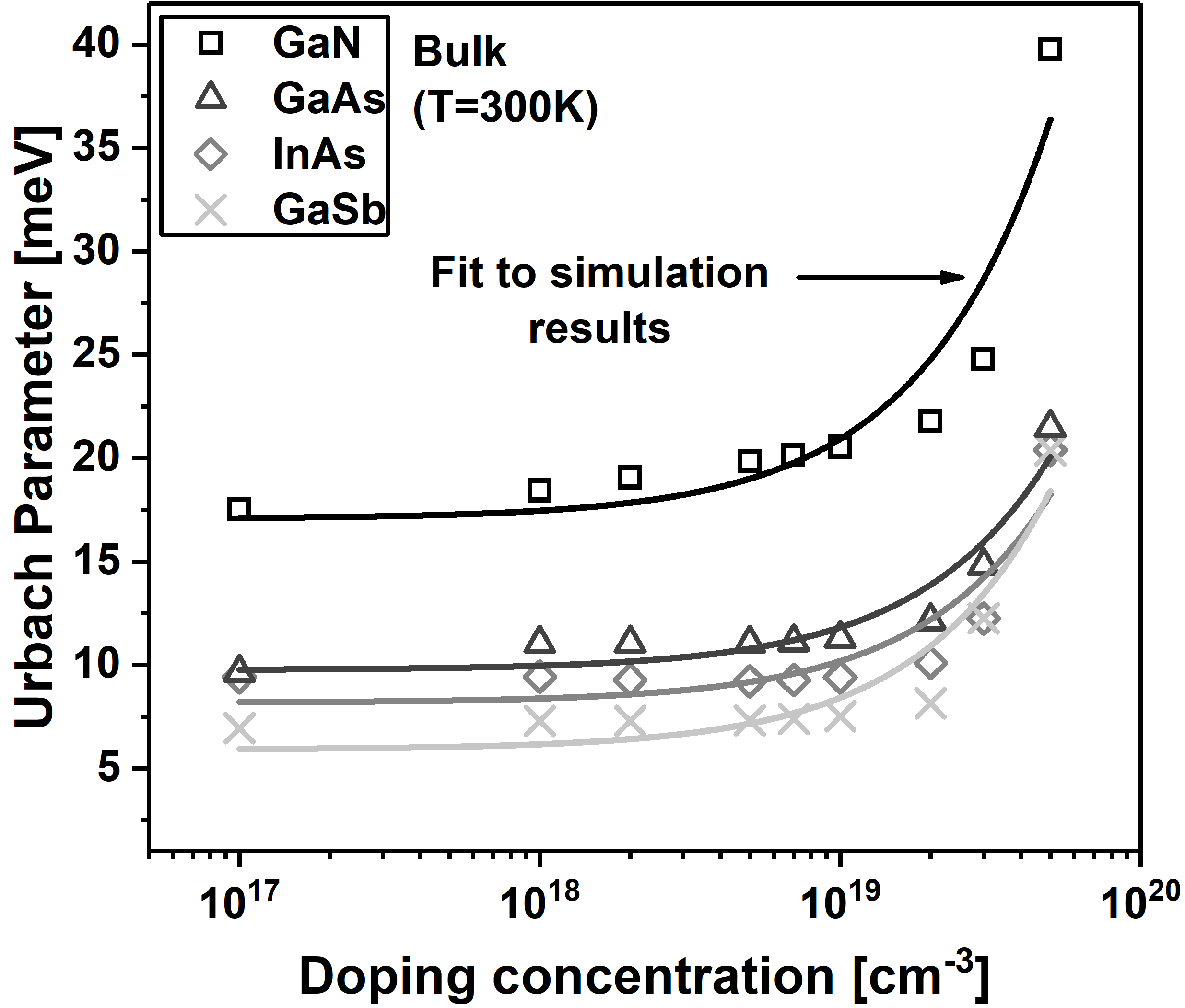}
\centering
\caption{Urbach parameter of bulk GaN, GaAs, InAs and GaSb as a function of the doping concentration. 
Symbols represent the NEGF results and lines depict the respective fitting curve. 
}
\label{fig:urbach_versus_doping_bulk_fit}
\end{figure}

Figure~\ref{fig:urbach_versus_doping_bulk_fit} shows the simulated Urbach parameter of GaN, GaAs, InAs and GaN as a function of the doping concentration.
The behavior of the Urbach parameter with increasing doping concentration can be fit to $U\left(N_D\right)=U_{intrinsic} + A\left(N_{D}/10^{18}\right)^u$ (lines in Fig.~\ref{fig:urbach_versus_doping_bulk_fit}). 
We follow e.g. Refs.~\onlinecite{jain1990band, jain1991simple, lee1995determination} in using a polynomial fit function for the Urbach parameters and band gap narrowing.
The fitted parameters are given in Table~\ref{table:urbach_fit}. 

\begin{figure}
\includegraphics[scale=0.35]{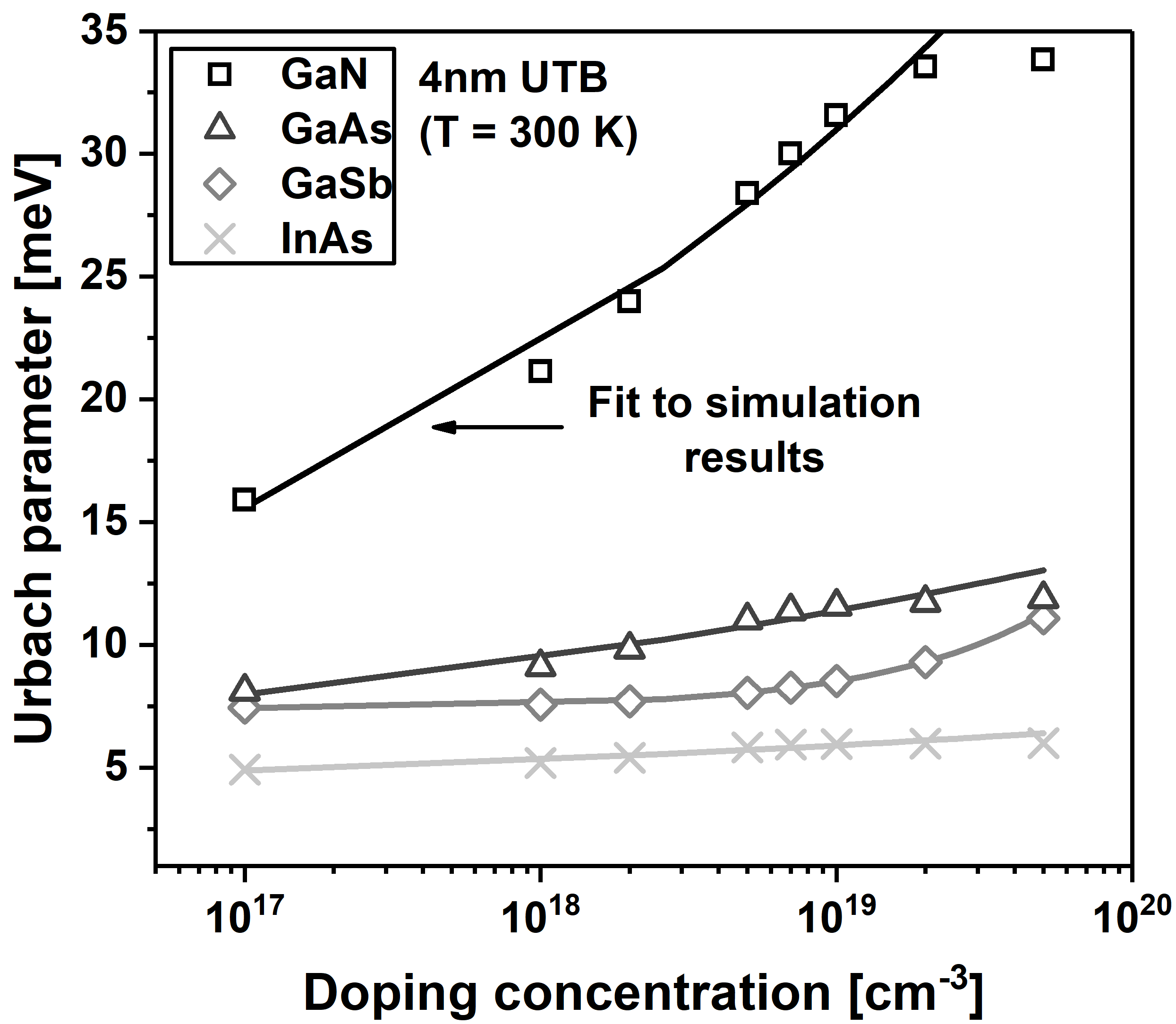}
\centering
\caption{Urbach parameter of 4nm thick GaN, GaAs, GaSb, and InAs UTBs as a function of the doping concentration. 
Symbols represent the NEGF results and lines depict the respective fitting curve.}
\label{fig:urbach_versus_doping_utb}
\end{figure}

\begin{figure}
\includegraphics[scale=0.35]{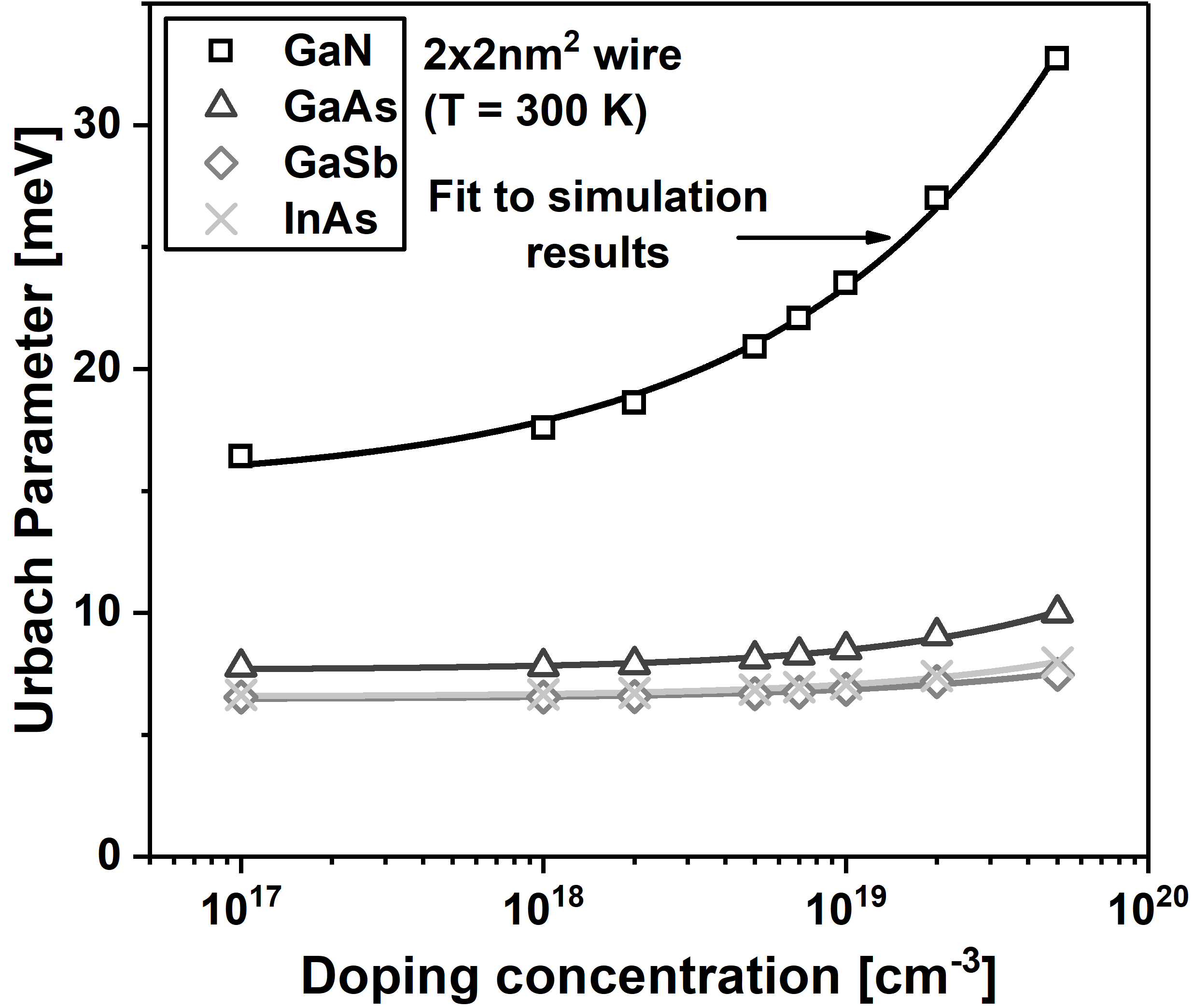}
\centering
\caption{Urbach parameter of $2\times2nm^2$ GaN, GaAs, GaSb, and InAs nanowires as a function of the doping concentration. 
Symbols represent the NEGF results and lines depict the respective fitting curve.}
\label{fig:urbach_versus_doping_wire}
\end{figure}

A similar dependence of the Urbach parameter on the doping concentration is found in Figs.~\ref{fig:urbach_versus_doping_utb} and \ref{fig:urbach_versus_doping_wire} for UTBs and nanowires, respectively. The fitted curves in Figs.~\ref{fig:urbach_versus_doping_utb} and \ref{fig:urbach_versus_doping_wire} capture the simulated results well. 
The corresponding fitting parameters are given in Table~\ref{table:urbach_fit}. The doping dependence of the Urbach parameter differs significantly for bulk, UTB and nanowires. 
For instance, it is approximately linear in the case of bulk and sub-linear with $u=2/3$ for nanowires. 
That behavior arises from three contributions - explicit doping dependence in impurity scattering potentials, implicit doping dependence through electrostatic screening and screening dependent non-local scattering ranges as can be seen from Eqs.~(\ref{eq:implessselfenergy1D}), (\ref{eq:implessselfenergy2D}), and (\ref{eq:implessselfenergy3D}). 
While the fitted exponent $u$ is uniform across materials for bulk and nanowires, it varies strongly for UTBs. 

\begin{table*}

\begin{centering}
\begin{ruledtabular}
\begin{tabular}{|c|c|c|c|c|}
\hline 
 & \textbf{GaN} & \textbf{GaAs} & \textbf{GaSb} & \textbf{InAs}\tabularnewline
\hline 
Bulk & $A=0.38,\; u=1$ & $A=0.20,\; u=1$ & $A=0.20,\; u=1$ & $A=0.25,\; u=1$\tabularnewline
\hline 
4nm UTB & $A=22.21,\; u=0.14$ & $A=2.04,\; u=0.10$ & $A=6.92,\; u=0.10$ & $A=0.18,\; u=0.77$\tabularnewline
\hline 
2$\times$2 $nm^{2}$ wire & $A=2.70,\; u=0.66$ & $A=0.17,\; u=0.66$ & $A=0.10,\; u=0.66$ & $A=0.11,\; u=0.66$\tabularnewline
\hline 
\end{tabular}
\end{ruledtabular}
\protect\caption{Parameters for variation of Urbach parameter with doping concentration for different materials for bulk, UTB and wire. Fitting for simulation results has been performed using the expression $U(N_D)= U_{intrinsic} + A(N_D/1E18)^u$}
\label{table:urbach_fit}
\par\end{centering}
\end{table*}

\subsection{Band gap narrowing predictions vs. temperature, doping and confinement}

\begin{figure}
\includegraphics[scale=0.35]{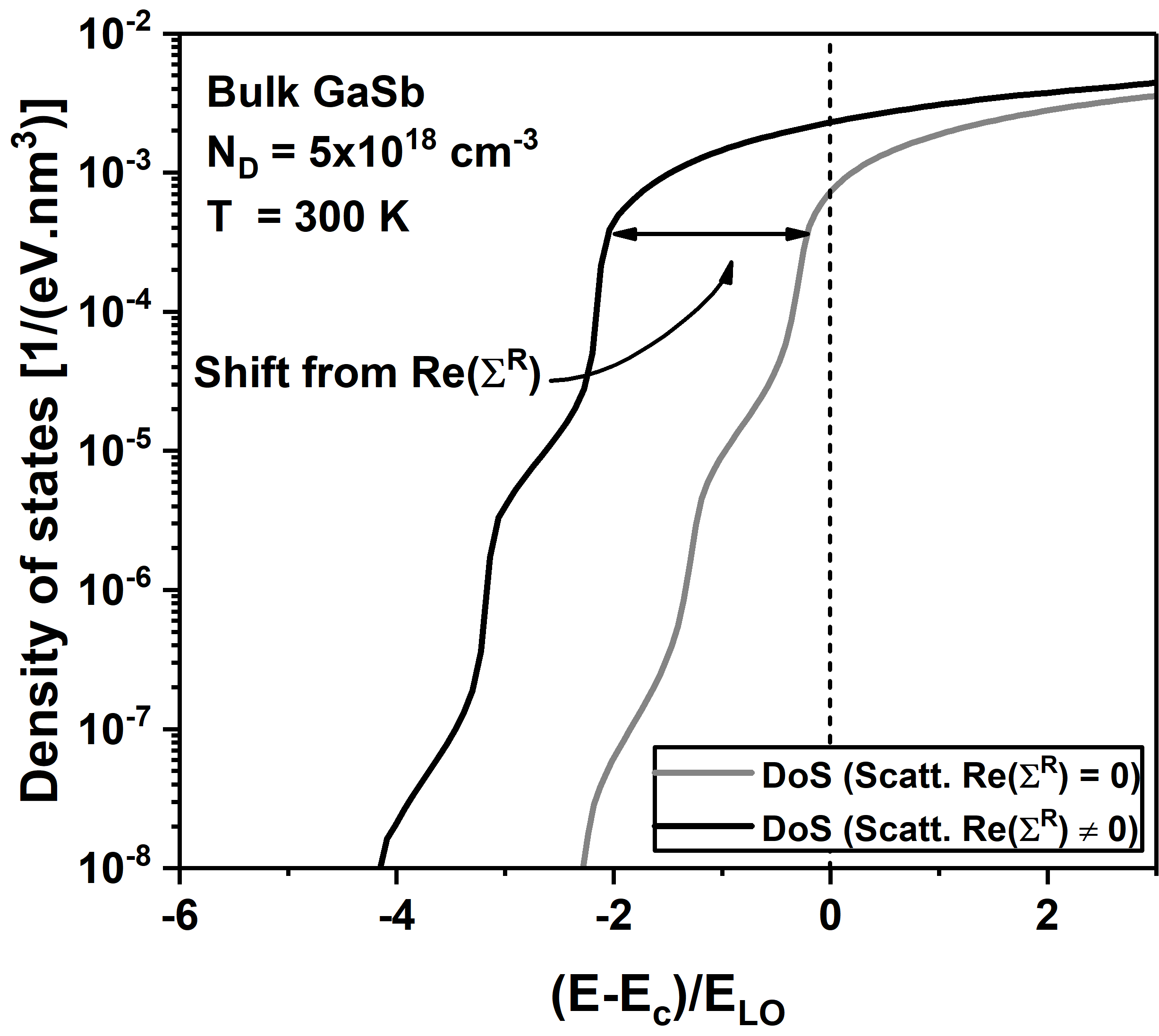}
\centering
\caption{Energy resolved conduction band density of states with and without the real part of the retarded pop and impurity scattering self-energies. 
The real part shifts the band edge to lower energies. 
The effective band gap narrowing equals the band edge difference caused by the real part of the retarded scattering self-energies.}
\label{fig:dos_with_and_without_real_sigma}
\end{figure}

As detailed in Sec.~\ref{section:urbach} and illustrated in Fig.~\ref{fig:dos_with_and_without_real_sigma}, the band gap narrowing is deduced from two simulations - one with the real parts of all retarded scattering self-energies neglected and one with real part fully included.
The band edge differences of the two simulations equals the band gap narrowing. 

To verify this approach, the NEGF predictions of band gap narrowing are compared against published experimental data. 
\begin{figure}
\includegraphics[scale=0.35]{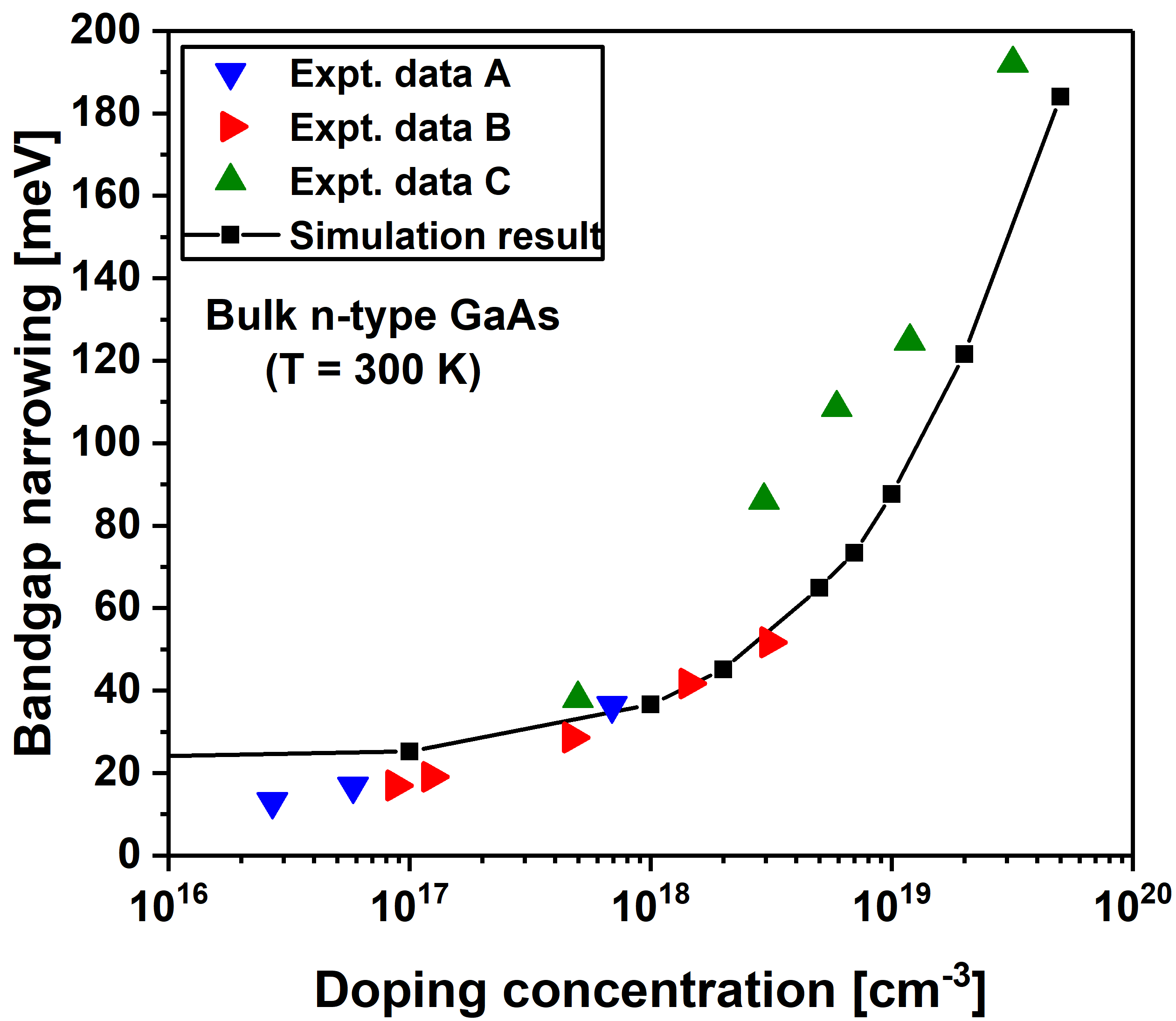}    
\centering
\caption{Band gap narrowing caused by conduction band edge shifts as a function of the n-type doping concentration of GaAs. 
NEGF results agree quantitatively with experimental results of Refs.~\onlinecite{yao1990plasmons} (labeled "Exp. data A"), \onlinecite{luo2002study} (labeled "Exp. data B") and \onlinecite{harmon1994effective} (labeled Exp. data C).}
\label{fig:bgn_versus_expt_GaAs}
\end{figure}
Figure~\ref{fig:bgn_versus_expt_GaAs} compares the simulated bandgap narrowing as a function of the doping concentration for bulk n-type GaAs with the experimental data of Refs.~\onlinecite{yao1990plasmons}, \onlinecite{luo2002study}, and \onlinecite{harmon1994effective}. 
The band gap narrowing increases with the doping concentration  due to increasing impurity scattering.
The simulation results show quantitative agreement with the experimental data for conduction band edge shift related band gap narrowing. 
The remaining deviations can originate from crystal defects, disorder and exciton interactions. 
Experiments show that valence band edge shifts narrow the band gap similarly as the conduction band. 
In contrast, the NEGF results of this work show only marginal changes of the valence band edges indicating that the approximation of a constant scalar screening length might be inappropriate for valence band changes (for further discussion please see Ref.~\cite{oschlies1992first}).

\begin{figure}
\includegraphics[scale=0.35]{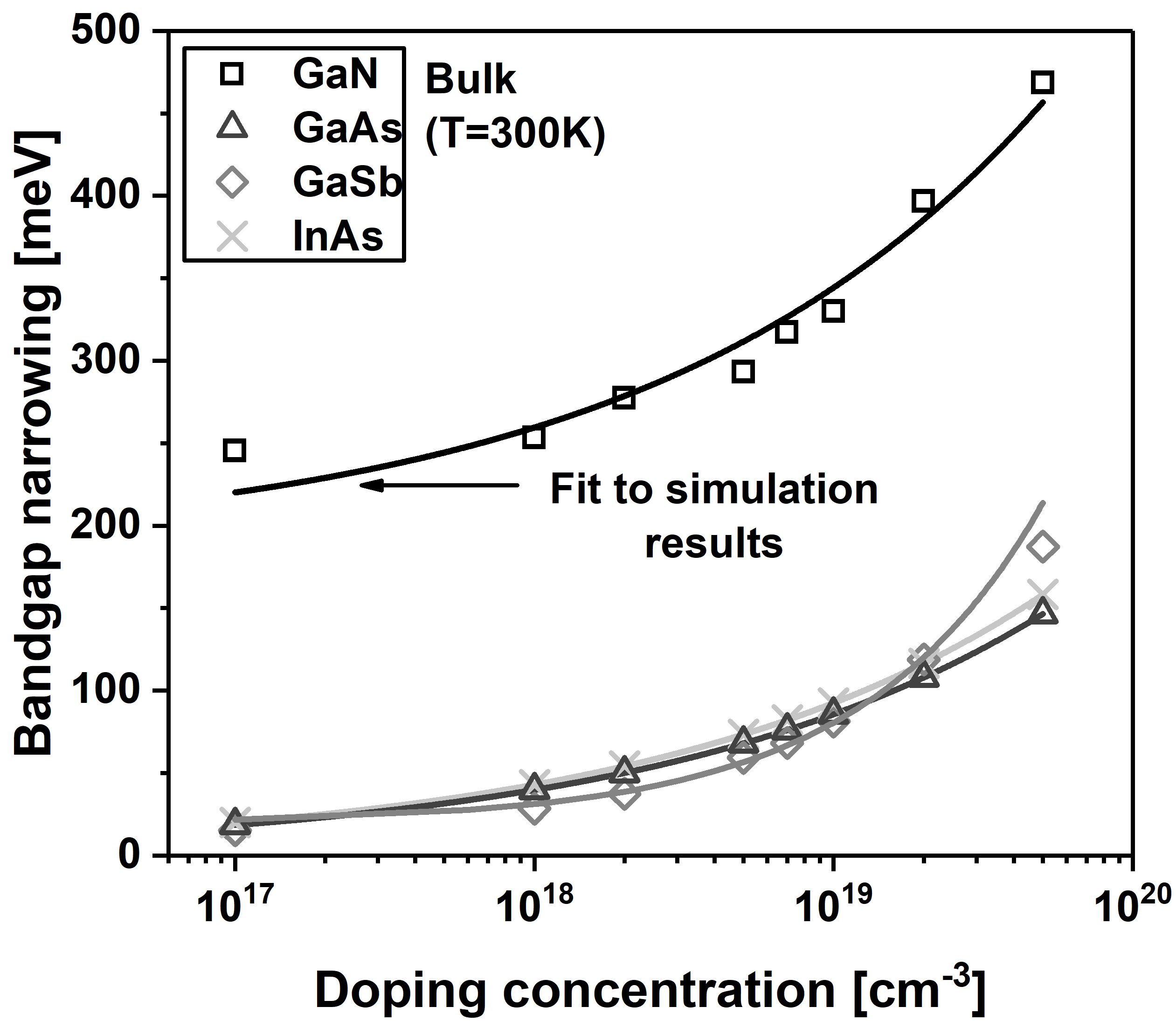}
\centering
\caption{Band gap narrowing caused by conduction band edge shifts as a function of the n-type doping concentration in bulk GaN, GaAs, GaSb, InAs and GaSb. 
Symbols correspond to the simulation results and lines correspond to fitted curves.}
\label{fig:bgn_versus_doping_fit}
\end{figure}

Figure~\ref{fig:bgn_versus_doping_fit} shows the conduction band induced band gap narrowing of GaN, GaAs, InAs and GaN as a function of the n-type doping concentration.
All materials follow similar trends. 
GaN, has a larger band gap response due to the large phonon energy and the $~3\times$ larger scattering potential compared to the other materials.
Similar to the Urbach parameter, the variation of the band gap narrowing with doping can be fit with $BGN\left(N_{D}\right)=BGN_{intrinsic} + B\left(N_{D}/10^{18}\right)^{v}$. 
Fig.~\ref{fig:urbach_versus_doping_bulk_fit}). 
We follow e.g. Refs.~\onlinecite{jain1990band, jain1991simple, lee1995determination} in using a polynomial fit function for the Urbach parameters and band gap narrowing.
The corresponding fit parameters are summarized in Table~\ref{table:bgn_list}.
\begin{figure}
\includegraphics[scale=0.35]{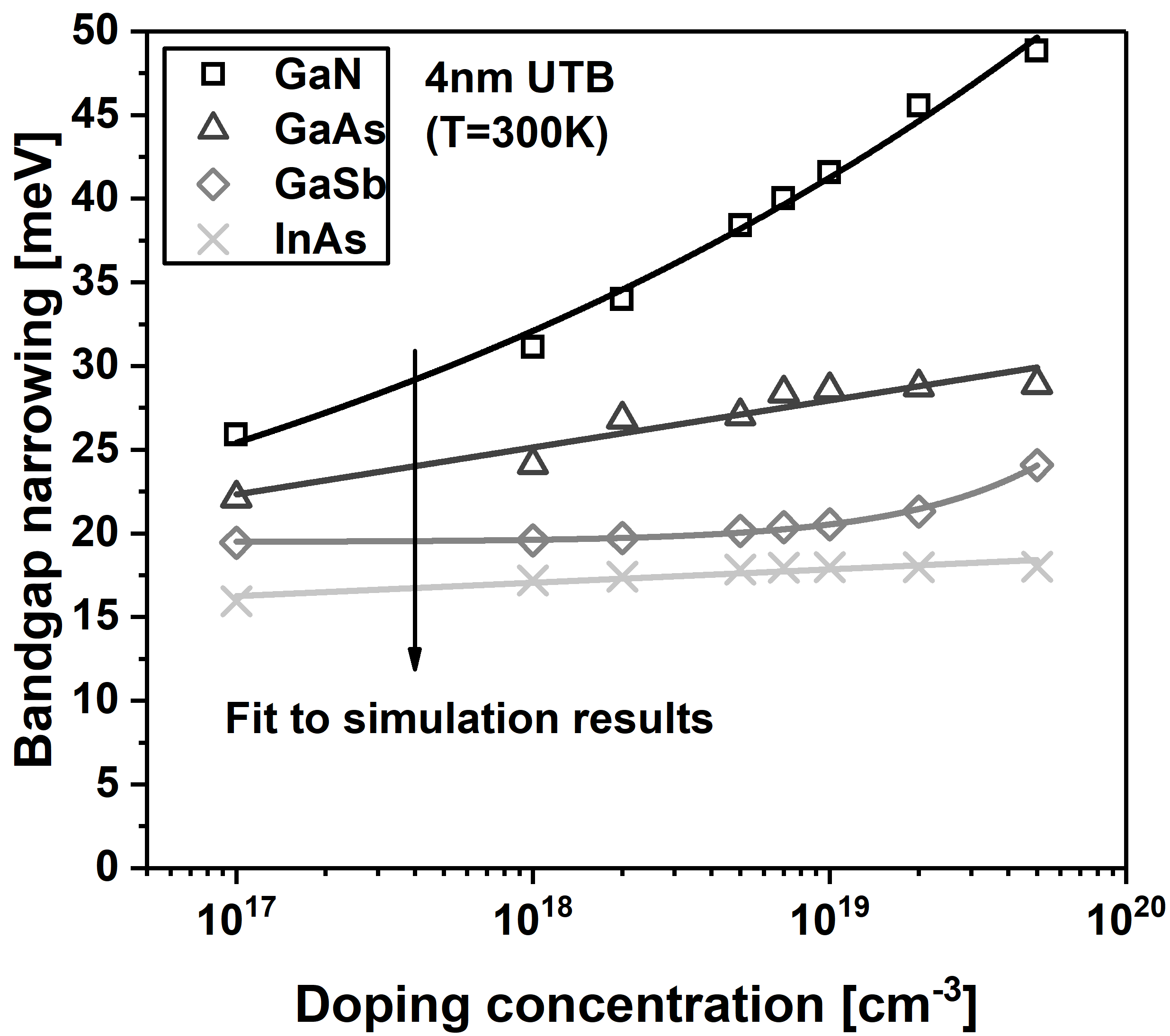}
\centering
\caption{Band gap narrowing caused by conduction band edge shifts as a function of the n-type doping concentration in 4nm thick GaN, GaAs, GaSb, InAs and GaSb UTBs. 
Symbols correspond to the simulation results and lines correspond to fitted curves.}
\label{fig:bgn_versus_utb_fit}
\end{figure}

\begin{figure}
\includegraphics[scale=0.35]{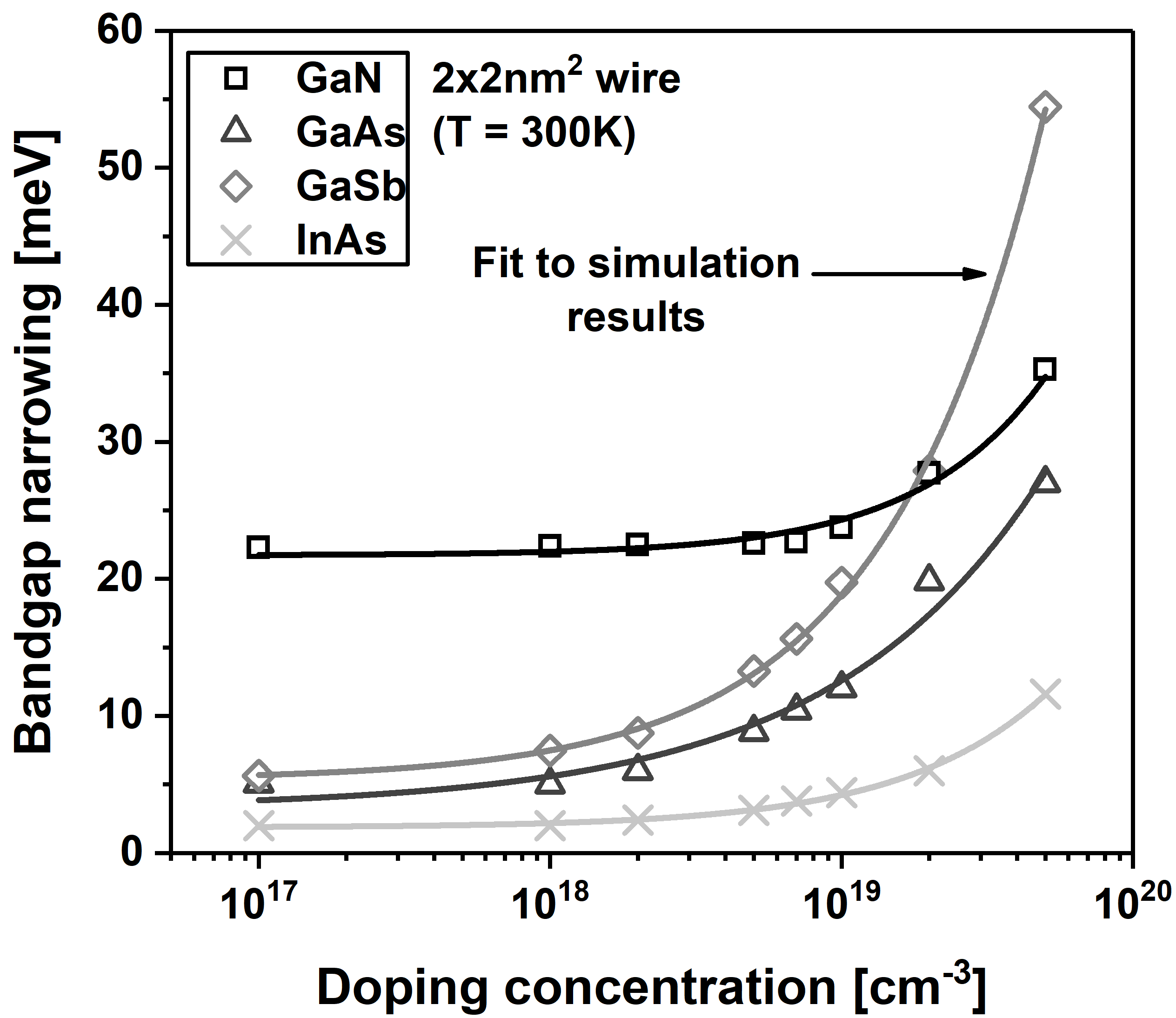}
\centering
\caption{Band gap narrowing caused by conduction band edge shifts as a function of the n-type doping concentration in a $2\times2~nm^{2}$ GaN, GaAs, GaSb, InAs and GaSb nanowire. 
Symbols correspond to the simulation results and lines correspond to the fitted curve.}
\label{fig:bgn_versus_wire_fit}
\end{figure}

The band gap narrowing for GaN, GaAs, InAs and GaN UTBs and nanowires as a function of the doping concentration are shown in Figs.~\ref{fig:bgn_versus_utb_fit} and \ref{fig:bgn_versus_wire_fit}.
Though the trend is similar in all three scenarios, the detailed doping dependence varies in bulk, UTBs and nanowires. 
This is similar to the scaling exponent $u$ of the Urbach parameter fit in Table~\ref{table:urbach_fit}.

\begin{table*}
\begin{centering}
\begin{ruledtabular}
\begin{tabular}{|c|c|c|c|c|}
\hline 
 & \textbf{GaN} & \textbf{GaAs} & \textbf{GaSb} & \textbf{InAs}\tabularnewline
\hline 
Bulk & $B=73.47,\; v=0.33$ & $B=53.80,\; v=0.33$ & $B=11.62,\; v=0.72$ & $B=39.57,\; v=0.33$\tabularnewline
\hline 
4nm UTB & $B=24.69,\; v=0.13$ & $B=15.6,\; v=0.12$ & $B=0.12,\; v=0.85$ & $B=0.18,\; v=0.77$\tabularnewline
\hline 
2$\times$2 $nm^{2}$ wire & $B=0.26,\; v=1$ & $B=2.13,\; v=0.8$ & $B=2.32,\; v=0.66$ & $B=0.31,\; v=0.87$\tabularnewline
\hline 
\end{tabular}
\end{ruledtabular}
\protect\caption{Parameters for variation of band gap narrowing parameter with doping concentration for different materials for bulk, UTB and wire. Fitting for simulation results has been performed using the expression $BGN(N_D)= BGN_{intrinsic} + B(N_D/1E18)^v$}
\label{table:bgn_list}
\par\end{centering}

\end{table*}

\section{Conclusion}

This work introduces self-energy formulas for bulk, nanowire, and ultra-thin body electrons scattering on 3D polar optical phonons and uniformly distributed charged impurities in the multi-band tight binding representation.
As often done, nonlocality of the scattering self-energies is limited to avoid unfeasible numerical load.
This underestimation of scattering is compensated with a scaling factor that is deduced from Fermi's golden rule prior to the actual NEGF calculations.
Band tails and band gap narrowing in GaAs, InAs, GaSb and GaN as well as in UTBs and nanowires of the same materials is predicted with NEGF in the self-consistent Born approximation. 
The extracted Urbach tail parameter as well as the conduction band driven band gap narrowing agrees quantitatively with available experimental data. 
This is also true for their dependence on the doping concentration.
A polynomial fit to the simulated Urbach tails and band gaps eases interpolating predictions from this work for bulk, UTBs and nanowires for different doping concentrations.

\section{Acknowledgements}
James Charles and Tillmann Kubis acknowledge support by the Silvaco Inc. 
This work was partly supported by the
Semiconductor Research Corporations Global Research Collaboration (GRC) (2653.001). This research was supported in part through computational resources provided by Information Technology at Purdue, West Lafayette, Indiana.
The authors acknowledge the Texas Advanced Computing Center (TACC) at The University of Texas at Austin for providing HPC resources that have contributed to the research results reported within this paper. 
This research used resources of the Oak Ridge Leadership Computing Facility, which is a DOE Office of Science User Facility supported under Contract DE-AC05-00OR22725. 

\bibliographystyle{apsrev4-1k.bst}
\bibliography{refs.bib}
  
\end{document}